\documentclass[journal]{IEEEtran}
\usepackage{color}
\usepackage[pdftex]{graphicx}
\usepackage{multicol}
\usepackage{times}
\usepackage{amssymb}
\usepackage{amsmath}
\usepackage{flafter}
\usepackage{indentfirst,latexsym,bm}
\usepackage{picinpar}
\usepackage{bm}
\usepackage{mathrsfs}
\usepackage{algorithm}
\usepackage[justification=centering]{caption}

\newtheorem{remark}{Remark}[section]
\newtheorem{theorem}{Theorem}[section]
\newtheorem{assumption}{Assumption}[section]
\newtheorem{lemma}{Lemma}[section]

\newtheorem{corollary}{Corollary}[section]
\newtheorem{example}{Example}[section]

\def\ban{\begin{eqnarray*}}
\def\ean{\end{eqnarray*}}
\def\bna{\begin{eqnarray}}
\def\ena{\end{eqnarray}}
\def\dref#1{(\ref{#1})}

\ifCLASSINFOpdf
\else
\fi
\begin{document}
%
\title{Convergence of the Distributed SG Algorithm Under Cooperative Excitation Condition}
%
%
%

\author{{Die Gan
       and Zhixin Liu.}
\thanks{Corresponding author: Zhixin Liu.}
\thanks{This work was supported by the National Key R$\&$D Program of China under Grant 2018YFA0703800.  Natural Science Foundation of China, NSFC 11688101.}
\thanks{D. Gan and Z. X. Liu are  with the  Key Laboratory of Systems and Control, Academy of Mathematics and Systems Science, Chinese Academy of Sciences,  and School of Mathematical Sciences, University of Chinese Academy of Sciences, Beijing, P. R. China. Emails: gandie@amss.ac.cn, lzx@amss.ac.cn.}
}

%
%

\markboth{IEEE Transactions on Neural Networks and Learning Systems,~Vol.~X, No.~X, XXX~2021}%
{Shell \MakeLowercase{\textit{et al.}}: Bare Demo of IEEEtran.cls for IEEE Journals}
%



\maketitle

\begin{abstract}
In this paper, a distributed stochastic gradient (SG) algorithm is proposed where the estimators are aimed to collectively estimate an unknown time-invariant parameter from a set of noisy measurements obtained by distributed sensors. The proposed distributed SG algorithm combines the consensus strategy of the estimation of neighbors with the diffusion of regression vectors.  A   cooperative excitation condition is introduced, under which the convergence of the distributed SG algorithm can be obtained without relying on the independency and stationarity assumptions of regression vectors which are commonly used in existing literature. Furthermore, the convergence rate of the algorithm can be established. Finally, we show that all sensors can cooperate to fulfill the estimation task even though any individual sensor can not by a simulation example.
\end{abstract}


%
\IEEEpeerreviewmaketitle

\section{Introduction}
%
%
%
%
\IEEEPARstart{P}{arameter} estimation or filtering is one of the important issues in diverse fields including statistical learning, signal processing, system identification and adaptive control. With the development of computer science and communication, sensor networks are widely applied due to the advantages of flexibility, fault tolerance, and ease of deployment. The sensor networks bring more and more data, and how to apply these data to design proper estimation algorithms  is a promising research direction.

Generally speaking, there are three manners to process the information from the sensors: centralized, distributed and a combination of both (cf., \cite{c9}). For the centralized method, the information measured by the sensors are transmitted to a fusion center which uses all information to estimate the unknown signals or parameters. Compared with the distributed algorithms, the centralized ones are lack of robustness in addition to the burden brought by a large amount of computation and communication. In distributed algorithms, the sensors can accomplish complicated tasks in a cooperative manner even though each sensor can only receive local information.  A amount of theoretical results on distributed estimation or learning algorithms (e.g.,  \cite{c18}\cite{cccc1}\cite{cccc2}) arise because of comprehensive practical applications in engineering systems, such as target localization, collaborative spectral sensing, see e.g., \cite{c10}\cite{e9}.

In the investigation of distributed estimation algorithms, how to use the local information to design the algorithms is important for the property of the algorithms. Three types of strategies are often adopted in the current literature: incremental strategy (cf., \cite{e7}), consensus strategy (cf., \cite{e5}), and diffusion strategy (cf., \cite{c16}\cite{e8}). Based on these three strategies, many different distributed adaptive estimation algorithms  are proposed, such as the diffusion least mean squares (LMS), the consensus-based Kalman filter, the diffusion least squares. Correspondingly, the stability and the convergence analysis of the distributed estimation algorithms are also investigated under some signal conditions (cf.,\cite{e1}-\cite{c7}).
Schizas et al.  in \cite{cc2}  established the stability results for a distributed LMS-type adaptive algorithm with the strictly stationary ergodic regressor vectors. Takahashi et al. in \cite{cc3} investigated the mean-transient and mean-square performance analysis of the diffusion LMS algorithm with the  independent and identically distributed regressors.
Cattivelli and Sayed in \cite{cc1} provided
the steady-state mean and mean-square analysis  of the diffusion LMS algorithm with the independent Gaussian regressors. Arablouei et al. in \cite{cc4} presented the convergence analysis of a partial-diffusion recursive least squares algorithm for the independent and ergodic input vectors. Lei and Chen in \cite{c7} studied the convergence of the distributed stochastic approximation algorithm with ergodic signals. So far,  most results require that the regression signals satisfy some stringent conditions, such as independency, stationarity and ergodicity assumptions, which makes it hard or even impossible to apply these theoretical results to practical feedback control systems. We remark that a preliminary attempt towards the relaxation of the independency and stationarity assumptions was made by Chen, Liu and Guo (cf., \cite{c13}), where they provided a cooperative excitation condition to guarantee the stability of the diffusion LMS algorithm, and some elegant results for the distributed LMS algorithm were further established by Xie and Guo in  \cite{c16} and \cite{c15}  under a general cooperative information condition.

It is well-known that the standard SG algorithm  has the advantages of simple expression and easy computation. The stochastic gradient algorithm is widely applied in the area of adaptive control, and also has deep connections with stochastic gradient descent algorithm and its variants which are widely used to deal with optimization problems in machine learning \cite{d1}.
With the  development of  sensor networks, the distributed implementations of stochastic gradient algorithms have  attracted much attention of researchers (cf., \cite{new8}-\cite{new14}). For example,
George et al. in  \cite{new12}  proposed a distributed stochastic gradient descent algorithm for solving non-convex optimization
problems with applications in distributed supervised learning.
Cavalcante and Stanczak  in \cite{new14} devised and analyzed a novel online distributed algorithm for dynamic optimization problems.

In this paper, we consider a network of sensors which are aimed
to collectively estimate an unknown time-invariant parameter. We propose a distributed SG algorithm based on the combined diffusion and consensus strategies and study the convergence properties of the proposed algorithm for a dynamic system.  The main contributions of this paper are summarized as follows.

($i$)  We propose a novel distributed SG algorithm where each sensor is only allowed to communicate with its neighbors. The information of the regression vectors are first diffused through the sensor networks, and then the estimation of the unknown parameters is obtained by using the consensus-based strategy.


($ii$) We introduce a  cooperative excitation condition on the regressor signals, under which the strong consistency of the distributed SG algorithm can be established. By the cooperative excitation condition, we see that the estimation task can be still fulfilled by the cooperation of multiple sensors even if any of them can not. The properties of the product of non-independent and non-stationary random matrices are investigated, which is crucial for our analysis.  Our results can be degenerated to the convergence results on the standard stochastic gradient algorithm (cf., \cite{c1}\cite{c5}).

($iii$) We finally establish the convergence rate of the distributed stochastic gradient algorithm under the cooperative excitation condition. We remark that our convergence results and the convergence rate of the proposed algorithm  are obtained without relying on the assumptions of the independency and stationarity of the regression signals, which makes it possible for applications to the stochastic feedback systems.

The rest of this paper is organized as follows. In Section \ref{section2}, we first propose the distributed SG algorithm, and introduce the cooperative excitation condition. A necessary and sufficient condition for the strong consistency of the proposed algorithm, and the conditions of the regressors for the convergence of the algorithm are given in Section \ref{convergence}. The convergence rate of the distributed SG algorithm is given in Section \ref{convergence rate}. A simulation example is given in Section \ref{simulation} to illustrate our theoretical results. The concluding remarks are made in the last section.

\section{Problem Formulation}\label{section2}

\subsection{Some Preliminaries}

In this paper, we use $\bm A\in\mathbb{R}^{m\times n}$ to denote an  $m\times n$-dimensional matrix. For a matrix $\bm A$, $\|\bm A\|$ denotes its Euclidean norm, i.e., $\|\bm A\|\triangleq(\lambda_{\max}(\bm A\bm A^T))^{\frac{1}{2}}$, where the notation $T$ denotes the transpose operator and $\lambda_{\max}(\cdot)$ denotes the largest eigenvalue of the matrix. The notations $\det(\cdot)$ and  $tr(\cdot)$ are  used to denote the determinant and trace of the corresponding matrix respectively. If all elements of a matrix are nonnegative, then it is a nonnegative matrix, and furthermore if   $\sum^n_{j=1}a_{ij}=1$ for all $i$, then it is called a stochastic matrix. The Kronecker product $\bm A\otimes \bm B$ of two matrices $\bm A=(a_{ij})\in \mathbb{R}^{m\times n}$ and $\bm B\in \mathbb{R}^{p\times q}$  is defined as
\begin{gather*}
\bm A\otimes \bm B=\left(
\begin{matrix}
a_{11}\bm B&\cdots &a_{1n}\bm B\\
\vdots&\ddots &\vdots\\
a_{n1}\bm B&\cdots &a_{nn}\bm B\\
\end{matrix}
\right)\in \mathbb{R}^{mp\times nq}.
\end{gather*}

Our purpose is to propose a distributed estimation algorithm based on the information from a set of sensors and investigate the convergence properties of the proposed algorithm.
The sensors in sensor networks are modeled as nodes, and the relationship between sensors are presented as an undirected weighted graph $\mathscr{G}=(\mathcal{V},\mathcal{E},\mathcal{A})$, where $\mathcal{V}=\{1,2,3,\cdots, n\}$ is the set of sensors (i.e., nodes), the edge set $\mathcal{E}\in \mathcal{V} \times\mathcal{V}$ denotes the communication between sensors, and $\mathcal{A}=(a_{ij})$ is the weighted matrix. The elements of the matrix $\mathcal{A}$ satisfy:  $a_{ij}>0$ if $(i, j)\in \mathcal{E}$ and $a_{ij}=0$ otherwise. The neighbor set of the sensor $i$ is denoted as $N^i=\{j\in \mathcal{V}, (i,j)\in\mathcal{E} \}$, and the sensor $i$ is also included in this set. Each sensor can only exchange information with its neighbors. A path of length $\ell$ is a sequence of nodes $\{i_1,...i_{\ell}\}$ satisfying $(i_j,i_{j+1})\in \mathcal{E}$ for all $1\leq j\leq \ell-1$. The graph $\mathscr{G}$ is called connected if for any two sensors $i$ and $j$, there is a path connecting them. The diameter $D(\mathscr{G})$ of the graph $\mathscr{G}$ is defined as the maximum length of paths between any two sensors. For simplicity of analysis, the properties of the distributed algorithm are considered under the condition that the weighted matrix $\mathcal{A}$ is symmetric and stochastic. Hence, the Laplacian matrix $\bm L$ of the graph $\mathscr{G}$ can be written as $\bm L=\bm I-\mathcal{A}$ with $\bm I$ being the identity matrix.

A classical result for the Laplacian matrix $\bm L$ can be stated as follows.
\begin{lemma}\rm\cite{c8}\label{ll1}
The Laplacian matrix $\bm L$ has at least one zero eigenvalue, with other eigenvalues positive and less than or equal to 2. Moreover, if the graph $\mathscr{G}$ is connected, then $\bm L$ has only one zero eigenvalue.
\end{lemma}

The following lemma is often used in our analysis and we list it as follows.
\begin{lemma}\cite{c2}\label{l2}
~Let~$D_t\triangleq 1+\sum^t_{j=1}d_j,~d_j\geq 0$,~then
\begin{gather}
\sum^{\infty}_{j=1}\frac{d_j}{D^{\alpha}_j}<\infty, ~~\forall~\alpha>1,\nonumber\\
\sum^{\infty}_{j=1}\frac{d_j}{D_j}=\infty~~{\rm iff} ~~\lim_{j\rightarrow\infty}D_j=\infty.\nonumber
\end{gather}
\end{lemma}

\subsection{Distributed SG Algorithm}

In this paper, we consider a network consisting of $n$ sensors. The signal model of each sensor  $i\in\{1,...,n\}$
is assumed to obey the following  time-invariant regression stochastic model,
\begin{align}
y^i_{k+1}=\bm\theta^T{\bm\varphi^i_k}+\varepsilon^i_{k+1},~~ k\geq0\label{5},
\end{align}
where $y^i_k$ is the scalar observation of the sensor $i$ at the time instant $k$, $\bm\varphi^i_k\in\mathbb{R}^m$ is the random regression vector which may be the function of the current and past inputs and outputs, $\bm\theta\in\mathbb{R}^m$ is an unknown  parameter to be estimated, and
$\{\varepsilon^i_{k}\}$ is a noise process.

We aim at designing a distributed adaptive estimation algorithm where all sensors cooperatively estimate the unknown parameter $\bm\theta$ of the stochastic dynamical system (\ref{5}) by using local information $\{y^j_{k+1},\bm\varphi^j_k\}_{j\in N^i}$, and further establishing the (almost sure) convergence property and the convergence rate of the proposed distributed algorithm.


The standard SG algorithm is commonly used in the area of adaptive control and system identification (e.g.,\cite{c5}\cite{c6}). Inspired by \cite{c18} and based on the standard SG algorithm in the context of stochastic adaptive control, we propose the following distributed SG algorithm to cooperatively estimate the unknown parameter $\bm\theta$. The detailed algorithm can be found in Algorithm \ref{table1}.
\begin{algorithm}\caption{Distributed SG Algorithm}\label{table1}
Given any initial estimate $\bm{\hat\theta}^i_0$  of each sensor $i$, the distributed SG algorithm is given as follows:

Step 1.~Set the initial value at each time instant $k$ for each sensor $i\in\{1,...,n\}$ as
\ban
 x^i_k(0)=\frac{\|\bm\varphi^i_k\|^2}{r^i_k}, ~~r^i_k\triangleq1+\sum_{j=1}^{k}{\left\|\bm\varphi^i_j\right\|}^2.
\ean
Step 2.~Perform the following diffusion process for $q=0,1, 2, \cdots, Q$
with $Q\geq D(\mathscr{G})$,
\ban
x^i_k(q+1)=\sum_{j\in N^i}a_{ij}x^j_k(q).
\ean
Step 3.~Update the estimate $\bm{\hat\theta}^i_{k}$ of the unknown parameter for each sensor $i$ as follows,
\bna
\bm z^i_k(\bm{\hat\theta}^i_k)&=&x^i_k(Q)\sum_{l\in N^i}a_{li}(\bm{\hat\theta}^i_k-\bm{\hat\theta}^l_k),\nonumber\\
\bm{\hat\theta}^i_{k+1}&=&\underbrace{\bm{\hat\theta}^i_k+\mu\frac{\bm\varphi^i_k}{r^i_k}(y^i_{k+1}-({\bm\varphi^i_k})^T\bm{\hat\theta}^i_k)}_{\rm { Standard ~SG}}\nonumber\\
&&\underbrace{-\mu\nu\sum_{j\in N^i}a_{ij}\left(\bm z^i_k(\bm{\hat\theta}^i_k)-\bm z^j_k(\bm{\hat\theta}^j_k)\right)}_{\rm{Consensus-based ~ item}},\label{6}
\ena
where $\mu$ and $\nu$ are two step sizes lying in $(0,1)$, $r^i_k$ is defined in Step $1$.
\end{algorithm}


\begin{remark}

Algorithm \ref{table1} designed by combining the consensus strategy of the estimation of neighbors with the diffusion strategy of regression vectors is online and updated from time to time by using new measurement data. We see that the right hand side of (\ref{6}) in Algorithm \ref{table1} consists of two parts: the
first part is the standard SG algorithm which tries to minimize the prediction
error by using the innovation, while the second part can be regarded as the result of minimizing the weighted distance between the estimates of the sensor $i$ and its neighbors.

\end{remark}

\begin{remark}
 In Step 2, the multi-step diffusion strategy of regression vectors is used, which was widely employed in the design of the distributed algorithms (e.g.,\cite{new1}-\cite{new3}). The  diffusion step $Q$
 plays an important role in establishing the contraction property of the product of random matrices   $\prod^k_{p=j}(\bm I_{mn}-\mu\bm G_p)$  (see the proof of Theorem \ref{t4} in Appendix \ref{tt4}).
 Moreover, by following the proof of Lemma \ref{l5}, we see that if the condition number of $\bm \Phi^T_k\bm \Phi_k$ is bounded, then the multi-step diffusion step can be removed.
\end{remark}

For convenience of analysis, we introduce the following notations (see Table \ref{biao1}). In Table \ref{biao1}, $col(\cdot,\cdots,\cdot)$ denotes the vector stacked by the specified vectors, and $diag(\cdot,\cdots,\cdot)$ denotes the block matrix formed in a diagonal manner of the corresponding vectors or matrices.
\begin{table}[!ht]
\caption{Some notations}\label{biao1}
\begin{center}
\begin{tabular}{|c|c|c|}
  \hline
  Notation & Definition & Dimension\\
  \hline
  $\bm Y_k$ & $\{y^1_k,\cdots,y^n_k\}$ & $1\times n$ \\
  $\bm \Phi_k$ & $diag\{\bm\varphi^1_k,\cdots,\bm\varphi^n_k\}$ & $mn\times n$ \\
  $\bm\Xi_k$&$\{\varepsilon^1_k,\cdots,\varepsilon^n_k\}$&$1\times n$\\
  $\bm\Theta$&$col\{\underbrace{\bm\theta,\cdots,\bm\theta}_{n}\}$&$mn\times1$\\
  $\hat{\bm\Theta}_k$&$col\{\hat{\bm\theta}^1_k,\cdots,\hat{\bm\theta}^n_k\}$&$mn\times1$\\
  $\widetilde{\bm\Theta}_k$&$col\{\widetilde{\bm\theta}^1_k,\cdots,\widetilde{\bm\theta}^n_k\}$,~ $\widetilde{\bm\theta}^i_k=\bm\theta-\hat{\bm\theta}^i_k$&$mn\times1$\\
  $\bm R_k$&$diag\{r^1_k,\cdots,r^n_k\}$&$n\times n$\\
  $\mathscr{L}$&$\bm L\otimes \bm I_m$,~~$\bm L$ is the Laplacian matrix&$mn\times mn$\\
  $\bm A_k$&${\bm \Phi_k}{{\bm{R}}^{-1}_k}{{\bm\Phi}^T_k}$&$mn\times mn$\\
  $\bm X_k(Q)$&$diag\{x^1_k(Q),\cdots,x^n_k(Q)\}$&$n\times n$\\
  $\bm G_k$&$\bm A_k+\nu\mathscr{L}(\bm X_k(Q)\otimes\bm I_m)\mathscr{L}$&$mn\times mn$\\
  \hline
\end{tabular}
\end{center}
\end{table}

By the notations in Table \ref{biao1}, we can rewrite (\ref{5}) and (\ref{6}) into the following matrix form,
\ban
\bm Y_{k+1}&=&\bm\Theta^T{{\bm\Phi}_k}+\bm\Xi_{k+1},\nonumber\\
\hat{\bm\Theta}_{k+1}&=&\hat{\bm\Theta}_k+\mu\bm\Phi_k{\bm{R}}^{-1}_k(\bm Y^T_{k+1}-{\bm\Phi^T_k}\hat{\bm\Theta}_k)\nonumber\\
&&-\mu\nu\mathscr{L}(\bm X_k(Q)\otimes\bm I_m)\mathscr{L}\hat{\bm\Theta}_k.
\label{9}
\ean
Let $\widetilde{\bm\Theta}_k=\bm\Theta-\hat{\bm\Theta}_k$. It is clear that $\mathscr{L}\bm\Theta=0$, then we have the following error equation,
\begin{align}
\widetilde{\bm\Theta}_{k+1}=(\bm I_{mn}-\mu\bm G_k)\widetilde{\bm\Theta}_k-\mu{\bm \Phi_k}{{\bm{R}}^{-1}_k}\bm\Xi^T_{k+1}.\label{10}
\end{align}

In order to proceed  with our analysis, we introduce some assumptions concerning the graph, regression vectors and the system noise.

\begin{assumption}\label{a1}
The undirected graph $\mathscr{G}$ is connected.
\end{assumption}

\begin{assumption}\label{a2}(Cooperative Excitation Condition) 
There exist two positive constants $N$ and $K_0$ such that for~$k\geq K_0$, the following
inequality is satisfied,
\begin{gather}\label{cooperative}
\frac{\lambda^{(k)}_{\max}}{\lambda^{(k)}_{\min}}\leq N\left(log(\|\bm R_k\|)\right)^{\frac{1}{3}}, {~~~\rm a.s.}
\end{gather}
where~$\lambda^{(k)}_{\max}$, $\lambda^{(k)}_{\min}$ represent the maximum and minimum eigenvalues of the matrix
$
\frac{n}{m}\bm I_m+\sum^n_{i=1}\sum^k_{j=1}\bm\varphi^i_j{\bm\varphi^i_j}^T,
$
and $\|\bm R_k\|\rightarrow\infty$ as $k\rightarrow\infty$.
\end{assumption}

\begin{remark}
We give an intuitive explanation for the necessity of the above cooperative excitation condition. Consider an extreme case where all regressor
vectors $\bm \varphi^i_j$ are equal to zero. It is clear that Assumption \ref{a2} is not satisfied, and the unknown parameter $\bm{\theta}$ can not be identified since the observations do not contain any information about the unknown parameter. In order
to estimate $\bm{\theta}$, we should impose some nonzero information (or excitation) conditions
on the regressor vectors $\bm \varphi^i_j$. The following persistent excitation (PE) condition for the single sensor case is commonly used in the literature (see e.g.,\cite{new4}-\cite{new5}).

$\mathbf{(PE):}$ $r^i_k\xrightarrow[]{k\rightarrow\infty}\infty$ and for all $k\geq0$, there exits a positive constant $c_1$ such that
\begin{align}
\frac{\lambda_{\max}\left(\sum^k_{j=1}\bm\varphi^i_j{\bm\varphi^i_j}^T\right)}
{\lambda_{\min}\left(\sum^k_{j=1}\bm\varphi^i_j{\bm\varphi^i_j}^T\right)}\leq {c_1}<\infty,\label{PE} ~{\rm{a.s..}}
\end{align}
 Therefore, a similar excitation condition is needed to guarantee the convergence of the distributed SG algorithm.
\end{remark}

\begin{remark}
It can be  verified that the independent and  identically distributed (i.i.d.) signals (by the strong law of large numbers) and the stationary ergodic signals (by the ergodic theorem) have the following property,

$\mathbf{(Ergodicity~ Property):}$ For any $i\in\{1,...,n\}$, the regressor vectors $\bm\varphi^i_j$ satisfy the ergodicity property, i.e., there exists a matrix $\bm H_i$ such that
\ban
\frac{1}{k}\sum^k_{j=1}\bm\varphi^i_j{\bm\varphi^i_j}^T~~\underrightarrow{k\rightarrow\infty}~~\bm H_i, ~{\rm a.s.}.
\ean
Furthermore, if $\sum^n_{i=1}\bm H_i$ is positive definite (cf.,  \cite{c7}), then the ergodicity property implies the PE condition in the multiple sensor case, i.e., ${\lambda^{(k)}_{\max}}/{\lambda^{(k)}_{\min}}\leq {c_2}$ with ${c_2}$ being a positive constant, which means that Assumption \ref{a2} is satisfied.

\end{remark}

\begin{remark}
Guo in \cite{c1}  proved that the convergence of the standard SG algorithm  under the following excitation condition,
\begin{align} \label{excitaion}
\frac{\lambda_{\max}\Big(\sum^k_{j=1}\bm\varphi^i_j{\bm\varphi^i_j}^T\Big)}{\lambda_{\min}\Big(\sum^k_{j=1}\bm\varphi^i_j{\bm\varphi^i_j}^T\Big)}\leq \tilde{N}\Big(\log r^i_k\Big)^{\frac{1}{3}}, ~{\rm a.s.},
\end{align}
where $r^i_k\rightarrow\infty $ as $k\rightarrow\infty.$
 It is clear that the condition (\ref{excitaion}) is weaker than
the PE condition (\ref{PE}) for the single sensor case, and Assumption $\ref{a2}$ can be degenerated  to the condition (\ref{excitaion}) when $n=1$.
\end{remark}

\begin{remark} The cooperative excitation condition (i.e., Assumption \ref{a2}) reflects the cooperative effect of multiple sensors. By Example \ref{e1} in Section \ref{simulation}, we can show that the estimation task can be still fulfilled by the cooperation of multiple sensors even if any of them can not.
\end{remark}

\begin{assumption}\label{a3}
We assume that the system noise $\{\varepsilon^i_k,i=1,...n, k\geq1\}$ is a martingale difference sequence, that is,  $E(\bm\Xi_{k+1}|\mathscr{F}_k)=0$ with  $\mathscr{F}_k=\sigma\{\bm\varphi^i_j,\varepsilon^i_j,i=1,...n,j\leq k\}$ and $E(\cdot| \cdot)$ being  the conditional mathematical
expectation, and there exist constants $c_0>0$ and $\varepsilon\in[0,1)$~(which may depend on $\omega$) such that~$E(\|\bm\Xi_{k+1}\|^2|\mathscr{F}_k)\leq c_0\|\bm R_k\|^{\varepsilon}$ a.s..
\end{assumption}


\section{Convergence of Distributed SG Algorithm}\label{convergence}
In this section, we will provide the convergence analysis of the proposed distributed SG algorithm.

Let the state transition  matrix ${\bm\Psi}(k,j)$ be recursively defined by
\begin{align}
{\bm\Psi}(k+1,j)=(\bm I_{mn}-\mu\bm G_k){\bm\Psi}(k,j),
~{\bm\Psi}(j,j)=\bm I_{mn}.\label{11}
\end{align}
From the  error equation (\ref{10}), we can see that the analysis of the
error $\widetilde{\bm\Theta}_{k+1}$ can be divided into two key steps:

(i) Analyzing properties of the product of random matrices ${\bm\Psi}(k,j)=\prod_{p=j}^{k-1}(\bm I_{mn}-\mu\bm G_p)$.

(ii) Analyzing the cumulative effect of noises.

In order to establish the properties of ${\bm\Psi}(k,j)$, we first analyze the properties of $\bm G_k$.
\begin{lemma}\label{l1}
Suppose that Assumption $\ref{a1}$ is satisfied. If~$\mu(1+4\nu)\leq1$, then we have
\begin{gather}
0\leq\mu\bm G_k\leq\bm I_{mn}.\notag
\end{gather}
\end{lemma}
\textbf{Proof.} By Step 2 in Algorithm \ref{table1}, we have
\bna
x^i_k(Q)=\sum^n_{j=1}a^{(Q)}_{ij}\frac{\|\bm\varphi^j_k\|^2}{r^j_k},\label{xx}
\ena
where $a^{(Q)}_{ij}$  is the $i$-th row, $j$-th column element of $\mathcal{A}^Q$. The matrix $\mathcal{A}$ is stochastic, so is the matrix $\mathcal{A}^Q$ for $Q\geq 1$. Hence we have for $i\in\{1,\cdots,n\}$,
 \ban
 x^i_k(Q)\leq \Big(\max_{1\leq j\leq n}\frac{\|\bm\varphi^j_k\|^2}{r^j_k}\Big)\sum^n_{j=1}a^{(Q)}_{ij}=\max_{1\leq j\leq n}\frac{\|\bm\varphi^j_k\|^2}{r^j_k}
 \leq \|\bm A_k\|.
 \ean
By the definition $\bm X_k(Q)$ in Table \ref{biao1}, we have $\|\bm X_k(Q)\otimes\bm I_m\|\leq \|\bm A_k\|$. By Lemma \ref{ll1}, it follows that
\bna
\|\mu \bm G_k\|\leq \mu(\|\bm A_k\|+4\nu \|\bm A_k\|)\leq \mu(1+4\nu)\leq 1.\label{1}
\ena
This completes the proof of the lemma. \hfill$\blacksquare$

By the definition of $\bm G_k$ in Table \ref{biao1}, we see that $\bm G_k$ is a nonnegative definite matrix. Thus, there exists a matrix sequence $\{\bm B_k, k\geq 0\}$, such that for all $k$ we have
\bna
\bm B^2_k=\mu\bm G_k.\label{bk}
\ena

 In the following, we will analyze the properties of ${\bm\Psi}(k,j)$.
\begin{lemma}\label{l4}
Assume that the step sizes $\mu$ and $\nu$ satisfy  $\mu(1+4\nu)\leq1$. Then for any $k\geq0$ the following inequality holds,
\begin{align*}
\sum^{k-1}_{j=0}\|\bm \Psi(k,j+1)\bm B_j\|^2\leq mn.
\end{align*}
\end{lemma}
\textbf{Proof.}
By the definition of the state transition matrix in (\ref{11}), we have  $\bm\Psi(k,k)=\bm I_{mn}$ and \ban {\bm\Psi}(k,j+1){\bm\Psi}(j+1,j)={\bm\Psi}(k,j).\ean  Then,
\bna
mn&=&tr\Big({\bm\Psi}(k,k)\bm\Psi^T(k,k)\Big)\nonumber\\
&\geq& tr\Bigg(\sum^{k-1}_{j=0}\Big[{\bm\Psi}(k,j+1){\bm\Psi^T}(k,j+1)\nonumber\\
&&-{\bm\Psi}(k,j){\bm\Psi^T}(k,j)\Big]\Bigg)\nonumber\\
&=& tr\Bigg(\sum^{k-1}_{j=0}{\bm\Psi}(k,j+1)\Big[\bm I_{mn}-{\bm\Psi}(j+1,j)\nonumber\\
&&\cdot{\bm\Psi^T}(j+1,j)\Big]{\bm\Psi^T}(k,j+1)\Bigg)\nonumber.
\ena
Furthermore, by ${\bm\Psi}(j+1,j)=\bm I_{mn}-\mu\bm G_j$, we have
\bna
mn&\geq&tr\Bigg(\sum^{k-1}_{j=0}{\bm\Psi}(k,j+1)\Big[\mu\bm G_j+\mu\bm G_j(\bm I_{mn}-\mu\bm G_j)\Big]\nonumber\\
&&\cdot{\bm\Psi^T}(k,j+1)\Bigg)\nonumber\\
&\geq& tr\Bigg(\sum^{k-1}_{j=0}{\bm\Psi}(k,j+1)\mu\bm G_j{\bm\Psi^T}(k,j+1)\Bigg)\label{35}\\
&=&tr\Bigg(\sum^{k-1}_{j=0}{\bm\Psi}(k,j+1)\bm B^2_j{\bm\Psi^T}(k,j+1)\Bigg)\nonumber\\
&\geq&\sum^{k-1}_{j=0}\|{\bm\Psi}(k,j+1)\bm B_j\|^2\nonumber,
\ena
which completes the proof of the lemma. \hfill$\blacksquare$

How to deal with the noise effect of the distributed SG algorithm is a crucial step for the convergence analysis of the algorithm.
The following lemma provides an upper bound of the cumulative summation of the noises.

\begin{lemma}\label{l3}
Suppose that Assumption $\ref{a3}$ is satisfied, and the condition number of $\bm R_k$ is bounded \big(i.e., there exists a positive constant $\gamma$ which may depend on the sample $\omega$ such that $\max\limits_{1\leq i\leq n} {r^i_k} \Big/{\min\limits_{1\leq i\leq n}{r^i_k}}$ $ \leq\gamma$\big), then~$\bm S_k$ tends to a finite limit $\bm S$ as~$k\rightarrow\infty$,  where  $\bm S_k\triangleq\sum^k_{j=0}{\bm \Phi_j}{{\bm{R}}^{-1}_j}\bm\Xi^T_{j+1}$. Furthermore, there exists a positive constant $c$ which may depend upon the sample $\omega$ such that
\begin{gather}
\left\|\widetilde{\bm S}_{k-1}\right\|\leq c\|\bm R_k\|^{-\delta},
\end{gather}
where $\widetilde{\bm S}_{k-1}\triangleq\bm S -\bm S_{k-1}$ and $\delta\in(0, \frac{1-\varepsilon}{2})$ with $\varepsilon$ defined in Assumption \ref{a3}.
\end{lemma}

The proof is put in Appendix \ref{l3z}.

Now, we present a necessary and sufficient condition for the strong consistency of the distributed SG algorithm.

\begin{theorem}\label{t1}
Suppose that the condition number of $\bm R_k$ is bounded, and $\mu(1+4\nu)<1$. Then under Assumptions $\ref{a1}$ and $\ref{a3}$, the estimate $\hat{\bm\Theta}_k$ defined in Table \ref{biao1} converges to the true parameter ${\bm\Theta}$ almost surely (a.s.) for any initial value $\hat{\bm\Theta}_0$ if and only if~$\bm \Psi(k,0)\rightarrow0$ almost surely as $k\rightarrow\infty$.
\end{theorem}
\textbf{Proof.} By $(\ref{10})$~and $(\ref{11})$, we have the following expression,
\bna
\widetilde{\bm\Theta}_{k+1}&=&{\bm\Psi}(k+1,0)\widetilde{\bm\Theta}_0\nonumber\\
&&-\mu\sum^k_{j=0}{\bm\Psi}(k+1,j+1)\bm\Phi_j\bm R^{-1}_j\bm\Xi^T_{j+1}.\label{46}
\ena
 We notice that the second term on the right-hand side of $(\ref{46})$ is independent of~$\widetilde{\bm\Theta}_0$. Thus,  $\widetilde{\bm\Theta}_{k+1}\rightarrow0$ for any~$\widetilde{\bm\Theta}_0$ implies ~${\bm\Psi}(k+1,0)\widetilde{\bm\Theta}_0\rightarrow0$ as $k\rightarrow\infty$, which means that ${\bm\Psi}(k+1,0)\rightarrow0$ as $k\rightarrow\infty$. This completes the proof of the necessity part of the theorem.

Now, let us move on to the sufficiency part.
It is clear that in order to prove the convergence of the algorithm, we just need to prove
\begin{align}
\sum^k_{j=0}{\bm\Psi}(k+1,j+1)\bm\Phi_j\bm R^{-1}_j\bm\Xi^T_{j+1}\rightarrow0\ {\rm a.s.},\  \hbox{as}~k\rightarrow\infty.\label{12}
\end{align}

Set $\bm S_{-1}=0$.
By the definition of ${\bm\Psi}(\cdot,\cdot)$ in (\ref{11}), we have
\bna
&&\bigg\|\sum^k_{j=0}{\bm\Psi}(k+1,j+1)\bm\Phi_j\bm R^{-1}_j\bm\Xi^T_{j+1}\bigg\|\nonumber\\
&=&\bigg\|\sum^k_{j=0}{\bm\Psi}(k+1,j+1)(\bm S_j-\bm S_{j-1})\bigg\| \nonumber\\
&=&\bigg\|\bm S_k-\sum^k_{j=0}[{\bm\Psi}(k+1,j+1)-{\bm\Psi}(k+1,j)]\bm S_{j-1}\bigg\| \nonumber\\
&=&\bigg\|\bm S_k-\sum^k_{j=0}[{\bm\Psi}(k+1,j+1)-{\bm\Psi}(k+1,j)]\bm S \nonumber\\
&&+\sum^k_{j=0}[{\bm\Psi}(k+1,j+1)-{\bm\Psi}(k+1,j)]\widetilde{\bm S}_{j-1}\bigg\| \nonumber\\
&=&\bigg\|\bm S_k-\bm S+{\bm\Psi}(k+1,0)\bm S+\sum^k_{j=0}{\bm\Psi}(k+1,j+1) \nonumber\\
&&\cdot[\bm I_{mn}-{\bm\Psi}(j+1,j)]\widetilde{\bm S}_{j-1}\bigg\|,\label{3.8}
\ena
where $\bm S, \bm S_k, \widetilde{\bm S}_{k-1}$ are defined in Lemma \ref{l3}.

By Lemma \ref{l3} we have $\bm S_k-\bm S\rightarrow0$. By the condition that  ${\bm\Psi}(k+1,0)\rightarrow0$ as $k\rightarrow\infty$, we have  ${\bm\Psi}(k+1,0)\bm S\rightarrow0$ as $k\rightarrow\infty$.
By Lemma \ref{l3} and H\"{o}lder inequality, we have
\bna
&&\bigg\|\sum^k_{j=0}{\bm\Psi}(k+1,j+1)[\bm I_{mn}-{\bm\Psi}(j+1,j)]\widetilde{\bm S}_{j-1}\bigg\|\nonumber\\
&=&\bigg\|\sum^k_{j=0}{\bm\Psi}(k+1,j+1)\mu\bm G_j\widetilde{\bm S}_{j-1}\bigg\|\notag\\
&\leq& c\sum^M_{j=0}\|{\bm\Psi}(k+1,j+1)\bm B_j\|\frac{\|\bm B_j\|}{\|\bm R_j\|^{\delta}}\nonumber\\
&&+c\Big(\sum^k_{j=M+1}\|{\bm\Psi}(k+1,j+1)\bm B_j\|^2\Big)^{\frac{1}{2}}\nonumber\\
&&~~~~\cdot\Big(\sum^k_{j=M+1}\frac{\|\bm B_j\|^2}{\|\bm R_j\|^{2\delta}}\Big)^{\frac{1}{2}}.\label{54}
\ena
Furthermore, by Lemma \ref{l2} and  Lemma \ref{l1} we have
\bna
&&\sum^{\infty}_{j=1}\frac{\|\bm B_j\|^2}{\|\bm R_j\|^{2\delta}}=\sum^{\infty}_{j=1}\frac{\|\mu\bm G_j\|}{\|\bm R_j\|^{2\delta}}\notag\\
&\leq&\mu(1+4\nu)\sum^{\infty}_{j=1}\frac{\|\bm A_j\|}{\|\bm R_j\|^{2\delta}}\notag\\
&\leq&\sum^{\infty}_{j=1}\frac{\|\bm \Phi_j\|^2
\|\bm R^{-1}_j\|}{\|\bm R_j\|^{2\delta}}
\leq\gamma\sum^{\infty}_{j=1}\frac{\|\bm \Phi_j\|^2}{\|\bm R_j\|^{1+2\delta}}<\infty.\label{55}
\ena
According to~$(\ref{55})$~and Lemma \ref{l4}, we see that the two terms on the right-hand-side of (\ref{54}) tend to zero if we first let~$k\rightarrow\infty,$~ and then let ~$M\rightarrow\infty$. Hence (\ref{12}) holds. This completes the proof of the theorem.
 \hfill$\blacksquare$

A key problem still remains unresolved:  what conditions on the regression signals $\{\bm \varphi^i_k\}$ can guarantee that $\bm\Psi(k,0)\rightarrow0$ as $k\rightarrow\infty$? In the following, we will prove that under the cooperative excitation condition (i.e., Assumption \ref{a2}) the convergence results for the distributed algorithm can be established.

Before stating the main theorem of this section, we first give a lemma which provides a key step for the convergence of the  matrix ${\bm\Psi}(k,0)$.

\begin{lemma}\label{l5}
Suppose that Assumption $\ref{a1}$ is satisfied. Then there exists a positive constant $\sigma$, such that the following inequality holds for all $ t\geq0$ and all $u\geq0$,
\begin{gather}
\lambda_{\min}\left(\sum^{\hat{g}(t+u)-1}_{k=\hat{g}(t)}\bm G_k\right)\geq\sigma\lambda_{\min}\left(\sum^{\hat{g}(t+u)-1}_{k=\hat{g}(t)}\sum^n_{i=1}\bm A^i_k\right),\label{neww7}
\end{gather}
where~$\bm A^i_k\triangleq\frac{\bm\varphi^i_k(\bm\varphi^i_k)^T}{r^i_k}$, $\hat{g}(t)\triangleq \max\{k:d_k\leq t\}$, and $d_k\triangleq\sum^n_{i=1}\sum^{k-1}_{j=K_0}\frac{\|\bm\varphi^i_j\|^2}{tr(\bm R_j)(\log tr(\bm R_{j-1}))^{\frac{1}{3}}}$, $K_0$ is defined in Assumption \ref{a2}.
\end{lemma}
\textbf{Proof.}
Set \ban & &\bm H^i_t=\sum^{\hat{g}(t+u)-1}_{k=\hat{g}(t)}\bm A^i_k,\ \ \
 \bm H_t= diag\{\bm H^1_t,...\bm H^n_t\},\\ & & \bm\Delta_t=\sum^{\hat{g}(t+u)-1}_{k=\hat{g}(t)}\bm G_k,\ \ \ \bm\Gamma_t=\sum^n_{i=1}\bm H^i_t.\ean
By the definition of $\bm A_k, \bm G_k$ in Table \ref{biao1}, we have
\begin{align*}
\bm\Delta_t&=\bm H_t+\nu\sum^{\hat{g}(t+u)-1}_{k=\hat{g}(t)}\mathscr{L}(\bm X_k(Q)\otimes\bm I_m)\mathscr{L},\\
\bm\Gamma_t&=\sum^{\hat{g}(t+u)-1}_{k=\hat{g}(t)}\sum^n_{i=1}\bm A^i_k.
\end{align*}

The eigenvalues of $\mathscr{L}$  are denoted in a non-decreasing order as $l_{1},\cdots,l_{m},l_{m+1},\cdots, l_{mn}$, and the corresponding unit orthogonal eigenvectors are denoted as $\xi_{1},\cdots,\xi_{m},\xi_{m+1},\cdots, \xi_{mn}$. By Assumption \ref{a1}, we see that $l_1=\cdots=l_m=0$ and
\begin{gather}
\bm \xi_1=\frac{1}{\sqrt n}\bm 1_n\otimes\bm e_1,...,\bm \xi_m=\frac{1}{\sqrt n}\bm 1_n\otimes\bm e_m, \label{xi}
\end{gather}
where $\bm 1_n$ denotes the $n-$diamensional vector with all entries equal to 1,  $\bm e_j ~(j=1,\cdots,m)$ is the $j$th column of the  identity matrix $\bm I_m$.

Hence for any unit vector $\bm \eta\in \mathbb{R}^{mn}$, we have the following expression,
\begin{gather*}
\bm \eta=\sum^m_{j=1}{\kappa_j}\bm\xi_j+\sum^{mn}_{j=m+1}{\kappa_j}\bm\xi_j\triangleq\bm \eta_1+\bm \eta_2,
\end{gather*}
where $\sum^m_{j=1}{\kappa^2_j}+\sum^{mn}_{j=m+1}{\kappa^2_j}=1$. Then we have
\bna
&&\bm\eta^T\bm \Delta_t\bm \eta \nonumber\\
&=&(\bm \eta_1+\bm \eta_2)^T\Big(\bm H_t+\nu\sum^{\hat{g}(t+u)-1}_{k=\hat{g}(t)}\mathscr{L}(\bm X_k(Q)\otimes\bm I_m)\mathscr{L}\Big)\nonumber\\
&&\cdot(\bm \eta_1+\bm \eta_2)\nonumber\\
&=&\bm\eta^T_1\bm H_t\bm\eta_1+\bm\eta^T_2\bm H_t\bm\eta_2+2\bm\eta^T_1\bm H_t\bm\eta_2\nonumber\\
&&+\nu\bm\eta^T_1\sum^{\hat{g}(t+u)-1}_{k=\hat{g}(t)}\mathscr{L}(\bm X_k(Q)\otimes\bm I_m)\mathscr{L}\bm\eta_1\nonumber\\
&&+\nu\bm\eta^T_2\sum^{\hat{g}(t+u)-1}_{k=\hat{g}(t)}\mathscr{L}(\bm X_k(Q)\otimes\bm I_m)\mathscr{L}\bm\eta_2\nonumber\\
&&+2\nu\bm\eta^T_1\sum^{\hat{g}(t+u)-1}_{k=\hat{g}(t)}\mathscr{L}(\bm X_k(Q)\otimes\bm I_m)\mathscr{L}\bm\eta_2\nonumber\\
&\triangleq &s_1+s_2+s_3+s_4+s_5+s_6.\label{s}
\ena

In the following, we will estimate $s_i (i=1, 2, \cdots, 6)$.
By the definition of $\bm\eta_1$, we see that $\bm\eta_1$ is the eigenvector corresponding to zero eigenvalue of $\mathscr{L}$. Hence we have $s_4=s_6=0$.

We note that $\bm H_t$ is a non-negative definite matrix. Thus we can  decompose  it as $\bm H_t=\bm H^{\frac{1}{2}}_t\bm H^{\frac{1}{2}}_t$,
 then
\ban
s_3&=&2\bm\eta^T_1\bm H_t\bm\eta_2
\geq-\zeta\bm\eta^T_1\bm H_t\bm\eta_1-\frac{1}{\zeta}\bm\eta^T_2\bm H_t\bm\eta_2\nonumber\\
&=&-\zeta s_1-\frac{1}{\zeta}s_2,\label{s3}
\ean
where the inequality $
2\bm M^T_1\bm M_2\leq\zeta\bm M^T_1\bm M_1+\frac{1}{\zeta}\bm M^T_2\bm M_2
$ is used, with $\zeta$ being a positive constant and $\bm M_1$ and $\bm M_2$ being two matrices with appropriate dimensions.

Let $y=\sum^m_{j=1}{\kappa^2_j}$. Then by (\ref{xx}) and the definition of $\bm X_k(Q)$, we have
\begin{align}
s_5&=\nu\bm\eta^T_2\sum^{\hat{g}(t+u)-1}_{k=\hat{g}(t)}\mathscr{L}(\bm X_k(Q)\otimes\bm I_m)\mathscr{L}\bm\eta_2\nonumber\\
&\geq \nu\sum^{\hat{g}(t+u)-1}_{k=\hat{g}(t)}\lambda_{\min}(\bm X_k(Q)\otimes\bm I_m)\sum^{mn}_{j=m+1}l^2_j{\kappa^2_j}\nonumber\\
&\geq a\nu\sum^{\hat{g}(t+u)-1}_{k=\hat{g}(t)}tr(\bm A_k)l^2_{m+1}(1-y)\nonumber\\
&=a\nu tr(\bm H_t)l^2_{m+1}(1-y),\label{s5}
\end{align}
where $a\triangleq\min\limits_{i,j\in\{1,\cdots,n\}}a^{(Q)}_{ij} $ is a positive constant for $Q\geq  D(\mathscr{G})$ (cf., \cite{c21}).

In the following, we will estimate $s_1$.
\bna
s_1&=&\bm\eta^T_1\bm H_t\bm\eta_1=(\sum^m_{j=1}{\kappa_j}\bm \xi_j)^T\bm H_t(\sum^m_{j=1}{\kappa_j}\bm \xi_j)\nonumber\\
&=&{\bm K}^T\bm \Xi^T\bm H_t\bm\Xi{\bm K},\label{ss1}
\ena
where ${\bm K}=({\kappa_1},...,{\kappa_j})^T$ and $\bm\Xi=(\bm \xi_1,...,\bm\xi_m)$. By the definition of $\bm \xi_i$  in (\ref{xi}), we have
\begin{align*}
\bm\Xi&=\frac{1}{\sqrt{n}}\left(
\begin{matrix}
\bm e_1&\bm e_2 &\cdots&\bm e_m\\
\bm e_1&\bm e_2 &\cdots&\bm e_m\\
\vdots&\vdots&\ddots&\vdots\\
\bm e_1&\bm e_2 &\cdots&\bm e_m\\
\end{matrix}
\right),\\
\bm H_t\bm\Xi&=\frac{1}{\sqrt{n}}\left(
\begin{matrix}
\bm H^1_t\bm e_1&\bm H^1_t\bm e_2 &\cdots&\bm H^1_t\bm e_m\\
\bm H^2_t\bm e_1&\bm H^2_t\bm e_2 &\cdots&\bm H^2_t\bm e_m\\
\vdots&\vdots&\ddots&\vdots\\
\bm H^n_t\bm e_1&\bm H^n_t\bm e_2 &\cdots&\bm H^n_t\bm e_m\\
\end{matrix}
\right),\\
&=\frac{1}{\sqrt{n}}(\bm H^1_t~\bm H^2_t\ldots\bm H^n_t)^T,\\
\bm\Xi^T\bm H_t\bm\Xi&=\frac{1}{n}\left(
\begin{matrix}
\bm e^T_1\bm H^1_t+e^T_1\bm H^2_t+\cdots+e^T_1\bm H^n_t\\
\bm e^T_2\bm H^1_t+e^T_2\bm H^2_t+\cdots+e^T_2\bm H^n_t\\
\vdots\\\bm e^T_m\bm H^1_t+e^T_m\bm H^2_t+\cdots+e^T_m\bm H^n_t\\
\end{matrix}
\right).\\
\end{align*}
Hence we have
$
\bm\Xi^T\bm H_t\bm\Xi=\frac{1}{n}\sum^n_{i=1}\bm H^i_t=\frac{1}{n}\bm\Gamma_t.
$
By this  and (\ref{ss1}) we have
\begin{gather}
s_1=\frac{1}{n}{\bm K}^T\bm\Gamma_t{\bm K}\geq\frac{\lambda_{\min}(\bm\Gamma_t)}{n}y. \label{s1}
\end{gather}
By the definition of $s_2$, we have
\begin{gather}
s_2=\bm\eta^T_2\bm H_t\bm\eta_2\leq tr(\bm H_t)(1-y). \label{s2}
\end{gather}
Substitute (\ref{s3})-(\ref{s2}) into (\ref{s}), we have for $\zeta\in(0,1)$,
\bna
\bm\eta^T\bm \Delta_t\bm \eta &\geq&(1-\zeta)s_1+(1-\frac{1}{\zeta})s_2+s_5\nonumber\\
&\geq&\frac{(1-\zeta)\lambda_{\min}(\bm \Gamma_t)}{n}y+(1-\frac{1}{\zeta})tr(\bm H_t)(1-y)\nonumber\\
&&+a\nu tr(\bm H_t) l^2_{m+1}(1-y).\label{80}
\ena
Thus, we have
\ban
\lambda_{\min}(\bm\Delta_t)&\geq&\Big[\frac{(1-\zeta)\lambda_{\min}(\bm \Gamma_t)}{n}-\Big(a\nu tr(\bm H_t) l^2_{m+1}\\
&&+tr(\bm H_t)-\frac{tr(\bm H_t)}{\zeta}\Big)\Big]y\\
&&+a\nu tr(\bm H_t) l^2_{m+1}+tr(\bm H_t)-\frac{tr(\bm H_t)}{\zeta}.
\ean
Taking $\zeta=\frac{1}{1+0.5l^2_{m+1}a\nu}\in(0,1)$,  then we can obtain the following inequality,
\ban
\lambda_{\min}(\bm\Delta_t)&\geq&\Big[\sigma\lambda_{\min}(\bm \Gamma_t)-0.5a\nu l^2_{m+1}tr(\bm H_t)\Big]y\\
&& +0.5a\nu l^2_{m+1}tr(\bm H_t),
\ean
where $\sigma\triangleq\frac{l^2_{m+1}a\nu}{2n+l^2_{m+1}a\nu n}\in(0,1).$
Hence by  $0\leq \lambda_{\min}(\bm \Gamma_t)\leq tr(\bm H_t)$ and $y\in(0,1)$, it is easy to obtain
\ban&&
\lambda_{\min}(\bm\Delta_t)- \sigma \lambda_{\min}(\bm \Gamma_t)\\&\geq& [\sigma \lambda_{\min}(\bm \Gamma_t)-0.5a\nu l^2_{m+1}tr(\bm H_t)](y-1)>0.
\ean
  This completes the  proof of the lemma. \hfill$\blacksquare$

\begin{remark} From Theorem \ref{t1}, we can see that
the properties of the product of the matrices $(I_{mn}-\mu\bm G_j), ~j\geq0$  is crucial for the strong consistency of the distributed SG algorithm. Lemma \ref{l5} establishes a connection between the
eigenvalues of $\{\bm A^i_k\}$ and $\{\bm G_k\}$, and thus builds a bridge between the
standard SG algorithm and the distributed SG algorithm.

\end{remark}

By using Lemma \ref{l5}, we have the following theorem.
\begin{theorem}
\label{t2}
Let $\mu(1+4\nu)<1$. Suppose that there exists $i_1\in\{1,\cdots,n\}$ such that~$\limsup_{k\rightarrow\infty}\frac{r^{i_1}_k}{r^{i_1}_{k-1}}\triangleq r^*<\infty$, and the condition number of $\bm R_k$ is bounded. Under Assumptions \ref{a1} and  \ref{a2}, we have $\bm \Psi(k,0)\rightarrow0$ as $k\rightarrow\infty$ a.s..
\end{theorem}

The proof of Theorem \ref{t2} is complicated, and we put it in Appendix \ref{t2z}.

\begin{remark}
For some typical cases such as the bounded sequence $\{\bm\varphi^i_k\} ~(i.e., ~c'\leq\|\bm \varphi^i_k\|\leq c'')$ and the i.i.d. sequence $\{\bm \varphi^i_k\}$, the condition $\limsup_{k\rightarrow\infty}\frac{r^{i_1}_k}{r^{i_1}_{k-1}}\triangleq r^*<\infty$  of Theorem \ref{t2} can be easily verified.
\end{remark}

By Theorems \ref{t1} and \ref{t2}, we have the following corollary.

\begin{corollary}\label{c3.1}
Under the conditions of Theorem \ref{t2}, if Assumption \ref{a3} is further satisfied, the convergence of the  distributed SG algorithm designed in Algorithm \ref{table1} can be obtained.
\end{corollary}
\textbf{Proof.} By Theorem \ref{t2}, we have $\bm \Psi(k,0)\rightarrow 0, $ as $k\rightarrow \infty$. Hence from Theorem \ref{t1}, the estimate $\hat{\bm\Theta}_k$ defined in  Table \ref{biao1} converges to the true parameter ${\bm\Theta}$ a.s., which completes the proof of the corollary. \hfill$\blacksquare$

\begin{remark} \label{re1}
The cooperative excitation condition proposed in Assumption \ref{a2} is reasonable in a sense that if there exists only one sensor which satisfies the excitation condition \dref {excitaion}, then all sensors in the network can satisfy Assumption \ref{a2}. We show this point under a mild condition (the condition  number of $\bm R_k$ is bounded) as follows,
\ban &&
\frac{{\lambda^{(k)}_{\max}}}{{\lambda^{(k)}_{\min}}}
\leq mn\gamma\cdot\frac{\lambda_{\max}\big\{\sum^k_{j=1}\bm\varphi^{i_2}_j{\bm\varphi^{i_2}_j}^T\big\}}{\lambda_{\min}\big\{\sum^k_{j=1}\bm\varphi^{i_2}_j{\bm\varphi^{i_2}_j}^T\big\}}\\
&\leq& mn\gamma \tilde{N}\Big(\log r^{i_2}_k\Big)^{\frac{1}{3}}
\leq mn\gamma \tilde{N}\left(\log(\|\bm R_k\|)\right)^{\frac{1}{3}}.
\ean
where ${i_2}$ denotes the index of the sensor satisfying
\dref {excitaion}, $\gamma$ is defined in Lemma \ref{l3}, ${\lambda^{(k)}_{\max}}$ and ${\lambda^{(k)}_{\min}}$ are defined in Assumption \ref{a2}.

Combining Remark \ref{re1} with Corollary \ref{c3.1},  we see that if only one sensor can fulfill the estimate task, then all sensors in our proposed algorithm can fulfill it.
\end{remark}

\begin{remark}
Different form most results in the literature, our results are obtained without using the independency and stationarity assumptions on the regression signals, which makes it possible to apply the distributed algorithm to practical feedback systems.
\end{remark}

\section{Convergence Rate of the Distributed SG Algorithm}\label{convergence rate}

In this section, we will consider the convergence rate of the distributed SG algorithm. In order to prove the theorems of this section, we first introduce the following two lemmas.

\begin{lemma}\label{l8}
If $\mu(1+4\nu)<1$,~then there exists a constant~$\tau_1\geq1$~such that for any~$k\geq0$, we have
\begin{align}
det(\bm I_{mn}-\mu\bm G_k)\geq [det(\bm I_{mn}-\bm A_k)]^{\tau_1}~.\nonumber
\end{align}
\end{lemma}
\textbf{Proof.} By the definition of $\bm G_k$ and \dref{1}, we have
\begin{align*}
det(\bm I_{mn}-\mu\bm G_k)&\geq(\lambda_{\min}(\bm I_{mn}-\mu\bm G_k))^{mn}\\
&=(1-\lambda_{\max}(\mu\bm G_k))^{mn}\\
&\geq[1-\lambda_{\max}({\mu(1+4\nu)}\bm A_k)]^{mn}\\
&=[\lambda_{\min}(\bm I-{\mu(1+4\nu)}\bm A_k)]^{mn}\\
&\geq\left[det(\bm I-{\mu(1+4\nu)}\bm A_k)\right]^{mn}\\
&\geq\left[det(\bm I-\bm A_k)\right]^{mn}.
\end{align*}
The lemma can be proved by taking $\tau_1=mn$. \hfill$\blacksquare$

\begin{lemma}\label{l9}
If~${\mu(1+4\nu)<1}$~,~then we have the following inequalities\footnotemark[1],
\ban
&(\lowercase\expandafter{\romannumeral1})& \|\bm \Psi(k,j)\|\leq1,~~~~0\leq j\leq k,~k\geq0;~\\
&(\lowercase\expandafter{\romannumeral2})& \frac{1}{\|\bm R_k\|^{\tau_1}}=O(\|{\bm\Psi}(k+1,0)\|^{m})~,~~~k\geq1~;\\
&(\lowercase\expandafter{\romannumeral3})& \|{\bm\Psi}(k,j+1)\|=O(\|{\bm\Psi}(k,0)\|\|\bm R_j\|^{n\tau_1});\\
&(\lowercase\expandafter{\romannumeral4})& \sum^{\infty}_{j={M}+1}\frac{\|\bm\Phi_j\|^2}{\|\bm R_j\|^{1+{\varsigma}}}\leq\frac{n^{1+{\varsigma}}}{{\varsigma}}\frac{1}{\|\bm R_M\|^{\varsigma}}~,{\varsigma}>0.
\ean
\footnotetext[1]{ Let $\{\bm A_k\}$  be a matrix sequence and $\{ b_k\}$  be a positive scalar
sequence. Then by $\bm A_k = O(b_k)$ we mean that there exists a constant $M > 0$ such that
$\| \bm A_k\|  \leq  M b_k, ~\forall k \geq  0$.}
where $\tau_1$ is defined in Lemma \ref{l8}.
\end{lemma}

The proof of the above lemma is given in Appendix \ref{s4}.

In the following, we establish the specific relationship between the error $\widetilde{\bm\Theta}_k$  and ${\bm\Psi}(k,0)$.
\begin{lemma}\label{t3}
Under the conditions of Theorem  \ref{t1}, if $\lim_{k\rightarrow\infty}\bm\Psi(k,0)=0$,  then we have the following inequality
\begin{gather}
\|\hat{\bm\Theta}_k- {\bm\Theta}\|=O\left(\|{\bm\Psi}(k,0)\|^{\frac{\delta}{n\tau_1 (1+\delta)}}\right) ~~{\rm a.s.} \nonumber,
\end{gather}
where $\delta$ is defined in Lemma \ref{l3} and  $\tau_1$  can be taken as $mn$.
\end{lemma}

\textbf{Proof.}
Let
\begin{align}\nonumber
&\beta(k)=\max\{j:\|\bm R_j\|^{n\tau_1}\leq k\},~~~~k\geq0,\\\nonumber
&\bar{\Delta}(k)=\beta(\|{\bm\Psi}(k,0)\|^{-\frac{1}{1+\delta}}),~~~k\geq0.
\end{align}
By the definition of~$\beta(k)$ and $\bar{\Delta}(k)$, we have
\begin{gather}\label{4.5}
\begin{aligned}[l]
&\|\bm R_{\bar{\Delta}(k)}\|^{n\tau_1}\leq \|{\bm\Psi}(k,0)\|^{-\frac{1}{1+\delta}},\\
&\|\bm R_{\bar{\Delta}(k)+1}\|^{n\tau_1}> \|{\bm\Psi}(k,0)\|^{-\frac{1}{1+\delta}}.
\end{aligned}
\end{gather}

According to Lemma $\ref{l9}\ (\lowercase\expandafter{\romannumeral3})$, we have
\begin{align}
\|{\bm\Psi}(k,\bar{\Delta}(k)+1)\|&=O(\|{\bm\Psi}(k,0)\|\cdot\|\bm R_{\bar{\Delta}(k)}\|^{n\tau_1})\nonumber\\
&=O(\|{\bm\Psi}(k,0)\|^{\frac{\delta}{1+\delta}}).\label{4.6}
\end{align}
We prove the following inequality by contradiction,
\bna
\bar{\Delta}(k)<k-1,\ \ \hbox{for large}\ \ k.\label{4.7}
\ena
Suppose that there exists a large constant $k_0$ such that $\bar{\Delta}(k_0)\geq k_0-1$. Then by \dref{4.5}, we have
\bna
\|\bm R_{k_0-1}\|^{n\tau_1}\leq \|{\bm\Psi}(k_0,0)\|^{-\frac{1}{1+\delta}}.\nonumber
\ena
By Lemma \ref{l9} $(ii)$, we  see that there exists a positive constant $\hat{c}$ such that
\bna
\|\bm R_{k_0-1}\|^{n\tau_1}\geq \hat{c}\|{\bm\Psi}(k_0,0)\|^{-mn}.\nonumber
\ena
Thus we have
\bna
\hat{c}\leq\|{\bm\Psi}(k_0,0)\|^{mn-\frac{1}{1+\delta}},\nonumber
\ena
which is  contradictory with ${\bm\Psi}(k,0)\rightarrow0$ as $k\rightarrow\infty$.

Note that by (\ref{1}), we have $\|\mu\bm G_j\|\leq\mu\frac{\|\bm \Phi_j\|^2}{\|\bm R_j\|}$.
Hence from Lemma \ref{l3} and \dref{3.8} in the proof of  Theorem $\ref{t1}$, we have the following estimation for the noise term of the system,
\ban
& &\left\|\sum^{k-1}_{j=0}{\bm\Psi}(k,j+1)\bm\Phi_j\bm R^{-1}_j\bm\Xi^T_{j+1}\right\|\nonumber\\
&\leq &\|\widetilde{\bm S}_{k-1}\|+\|{\bm\Psi}(k,0)\bm S\|\nonumber\\
& &+\sum^{k-1}_{j=0}\|{\bm\Psi}(k,j+1)\|\cdot\|\mu\bm G_j\|\cdot\|\widetilde{\bm S}_{j-1}\|\nonumber\\
&=&O(\|\bm R_k\|^{-\delta})+O(\|{\bm\Psi}(k,0)\|)\nonumber\\
& &+c\mu(1+4\nu)\gamma\sum^{k-1}_{j=0}\|{\bm\Psi}(k,j+1)\|\cdot\frac{\|\bm \Phi_j\|^2}{\|\bm R_j\|^{1+\delta}}.\nonumber\\
\ean

Combining this with (\ref{4.7}), we have
\bna
& &\left\|\sum^{k-1}_{j=0}{\bm\Psi}(k,j+1)\bm\Phi_j\bm R^{-1}_j\bm\Xi^T_{j+1}\right\|\nonumber\\
&=&O(\|\bm R_{\bar{\Delta}(k)+1}\|^{-\delta})+O(\|{\bm\Psi}(k,0)\|)\nonumber\\
& &+O\Big(\sum^{k-1}_{j=0}\|{\bm\Psi}(k,j+1)\|\cdot\frac{\|\bm \Phi_j\|^2}{\|\bm R_j\|^{1+\delta}}\Big).\label{140}
\ena

In the following, we estimate the last term on the right hand of (\ref{140}).
By Lemma \ref{l9} $(i)$ and \dref{4.6}, we have
\bna
& &\sum^{k-1}_{j=0}\|{\bm\Psi}(k,j+1)\|\cdot\frac{\|\bm \Phi_j\|^2}{\|\bm R_j\|^{1+\delta}}\nonumber\\
&\leq &\sum^{\bar{\Delta}(k)}_{j=0}\|{\bm\Psi}(k,\bar{\Delta}(k)+1)\|\cdot\|{\bm\Psi}(\bar{\Delta}(k)+1,j+1)\|\nonumber\\
& &\cdot\frac{\|\bm \Phi_j\|^2}{\|\bm R_j\|^{1+\delta}}+\sum^{k-1}_{j=\bar{\Delta}(k)+1}\|{\bm\Psi}(k,j+1)\|\cdot\frac{\|\bm \Phi_j\|^2}{\|\bm R_j\|^{1+\delta}}\nonumber\\
&=&O(\|{\bm\Psi}(k,0)\|^{\frac{\delta}{1+\delta}})\sum^{\infty}_{j=0}\frac{\|\bm \Phi_j\|^2}{\|\bm R_j\|^{1+\delta}}\nonumber\\
& &+\sum^{k-1}_{j=\bar{\Delta}(k)+1}\frac{\|\bm \Phi_j\|^2}{\|\bm R_j\|^{1+\delta}}.\label{4.8}
\ena

Furthermore,  by Lemma \ref{l2} and Lemma \ref{l9} $(iv)$, we have
\bna
& &\sum^{k-1}_{j=0}\|{\bm\Psi}(k,j+1)\|\cdot\frac{\|\bm \Phi_j\|^2}{\|\bm R_j\|^{1+\delta}}\nonumber\\
&= & O(\|{\bm\Psi}(k,0)\|^{\frac{\delta}{1+\delta}})+\frac{\|\bm \Phi_{\bar{\Delta}(k)+1}\|^2}{\|\bm R_{\bar{\Delta}(k)+1}\|^{1+\delta}}\nonumber\\
& &+\sum^{\infty}_{j=\bar{\Delta}(k)+2}\frac{\|\bm \Phi_j\|^2}{\|\bm R_j\|^{1+\delta}}\nonumber\\
&=&O(\|{\bm\Psi}(k,0)\|^{\frac{\delta}{1+\delta}})+O(\|\bm R_{\bar{\Delta}(k)+1}\|^{-\delta}).\nonumber
\ena
Combining the above inequality with (\ref{4.5}) and (\ref{140}), then we have
\bna
& &\Bigg\|\sum^{k-1}_{j=0}{\bm\Psi}(k,j+1)\bm\Phi_j\bm R^{-1}_j\bm\Xi^T_{j+1}\Bigg\|\nonumber\\
&=&O(\|\bm R_{\bar{\Delta}(k)+1}\|^{-\delta})+O(\|{\bm\Psi}(k,0)\|)\nonumber\\
& &+O(\|{\bm\Psi}(k,0)\|^{\frac{\delta}{1+\delta}})\nonumber\\
&=&O(\|{\bm\Psi}(k,0)\|^{\frac{\delta}{n\tau_1(1+\delta)}})+O(\|{\bm\Psi}(k,0)\|)\nonumber\\
& &+O(\|{\bm\Psi}(k,0)\|^{\frac{\delta}{1+\delta}})\nonumber\\
&=&O(\|{\bm\Psi}(k,0)\|^{\frac{\delta}{n\tau_1(1+\delta)}}),\label{141}
\ena
where $\|{\bm\Psi}(k,0)\|\leq1$ is used in the above inequality.
Hence by (\ref{46}) and \dref{141}, we have
\begin{gather}
\|\widetilde{\bm\Theta}_{k}\|=O(\|{\bm\Psi}(k,0)\|^{\frac{\delta}{n\tau_1 (1+\delta)}}) ~~{\rm a.s. }
\end{gather}
This completes the proof of the theorem.
 \hfill$\blacksquare$

We further establish the convergence rate under the cooperative excitation condition (Assumption \ref{a2}). We see that the rate can be expressed via simply characterizable quantities.

\begin{theorem}\label{t4}
Under the conditions of Theorem \ref{t2} and Assumption \ref{a3}, we have
\begin{gather*}
\|\widetilde{\bm\Theta}_{k}\|=O\Big((\log\|\bm R_k\|)^{-\delta_1}\Big) {~~\rm a.s. },
\end{gather*}
with $\delta_1$ being a positive constant.
\end{theorem}

The proof of the above theorem is given in Appendix \ref{tt4}.

\begin{remark}\label{nneww1}
By Theorem \ref{t4}, the convergence rate of the proposed distributed SG algorithm is mainly determined by the parameter $\delta_1$ and  $\|\bm R_k\|$. After a careful calculation by following the proof of  Theorem \ref{t4} in Appendix \ref{tt4}, we can obtain that the parameter $\delta_1$ is positively related to $\mu,\nu$ and $l_{m+1}$, where $l_{m+1}$ is the $(m+1)$-th eigenvalue of the Laplacian matrix $\mathscr{L}=\bm L\otimes \bm I_m$ in a non-decreasing order and can be regarded as a measure of the connectivity of the graph $\mathscr{G}$.
\end{remark}

\section{ A Simulation Example }\label{simulation}
In this section, we provide an example of non-independent regression signals to  illustrate the cooperative effect   in this paper.
\begin{example}\label{e1}
The network is composed of $n=28$ sensors whose dynamics obey the equation (\ref{5}) with $m=10$.  The system noises
$\varepsilon^i_k, (i=1,\cdots,n,~k\geq1)$  in (\ref{5})  are independent and identically distributed with $\varepsilon^i_k
\sim  \mathcal{N}(0, 1.2^2)$ (Gaussian distribution with zero mean and variance $1.2^2$). Set the unknown parameter as $\bm\theta=[1,2,\cdots,10]^T$.
Let $\bm \varphi^i_k\in\mathbb{R}^{10} ~(i = 1,\cdots,n, ~k\geq1)$ be generated by a state space model,
\bna
\bm u^i_k&=& \bm A_i \bm u^i_{k-1}+\bm B_i \xi^i_{k},\nonumber\\
\bm\varphi^i_k&=& \bm C_i \bm u^i_{k}.\label{phi}
\ena
For convenience, we take
\ban
\bm A_i&=&diag\{1.2,\cdots,1.2\} \in\mathbb{R}^{10\times10}, \\
\bm B_i&=&\bm {e_j}\in\mathbb{R}^{10},\\
\bm C_i&=&
col\{0,\cdots,0,\underset{j^{th}}{\bm {e_j}},0,\cdots,0\}\in\mathbb{R}^{10\times10},\\
\ean
where $j=\mod(i,m)$ and $\bm e_j ~(j=1,\cdots,m)$ is the $j$th column of the  identity matrix $\bm I_m$ ~$(m=10)$.

and let $\{ \xi^i_k, k \geq  1, i =1,\cdots, n\}$ be independent and identically distributed with $\xi^i_k\in\mathbb{R}
\sim  \mathcal{N}(0, 0.3^2)$.

The network structure is shown in Fig. \ref{network}. Here we use the Metropolis rule in \cite{new05} to construct
the weights.

\begin{figure}[!t]
\centering
\includegraphics[width=2.4in]{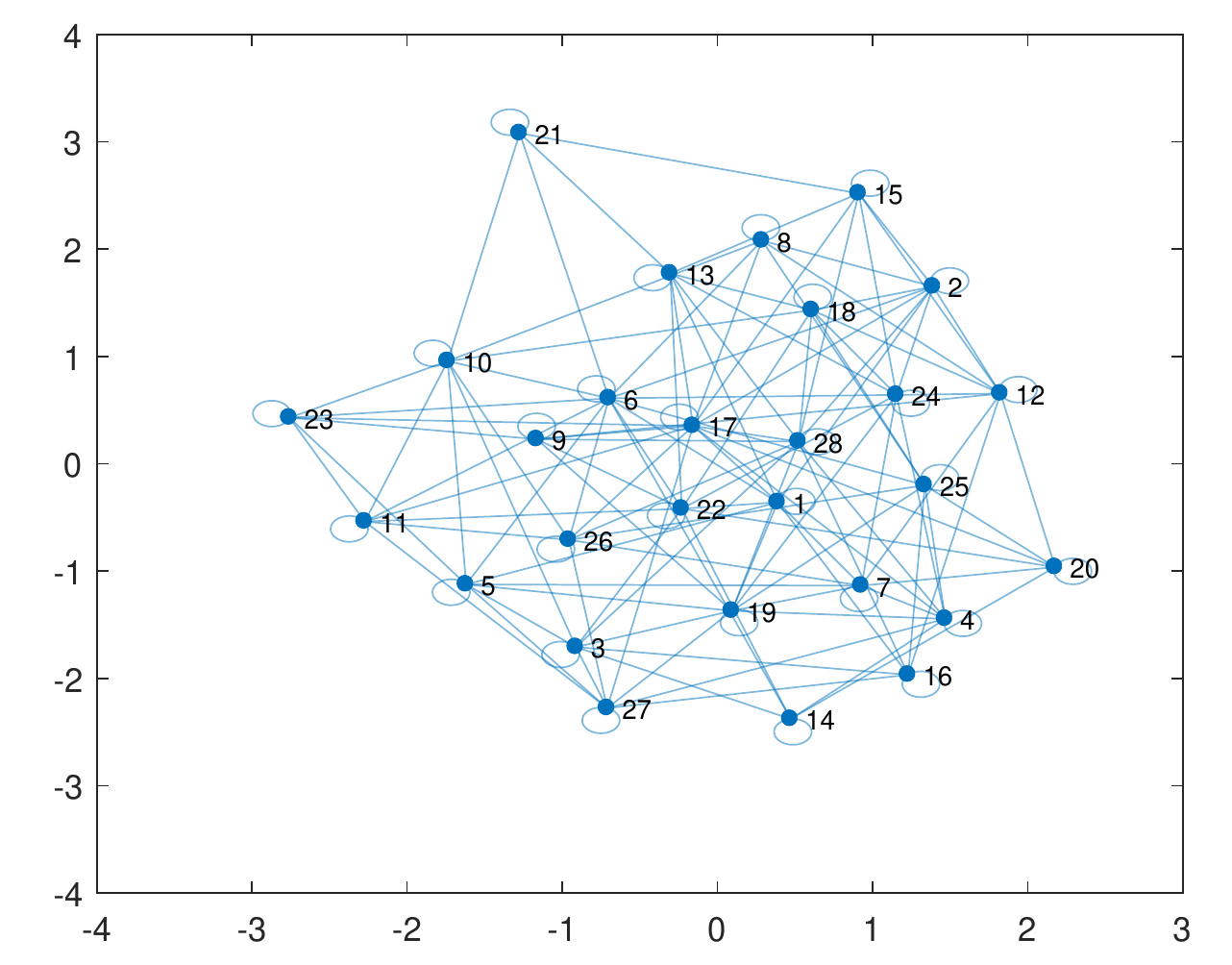}
\caption{Network topology}\label{network}
\end{figure}
\end{example}

\begin{figure}[!t]
\centering
\includegraphics[width=3in]{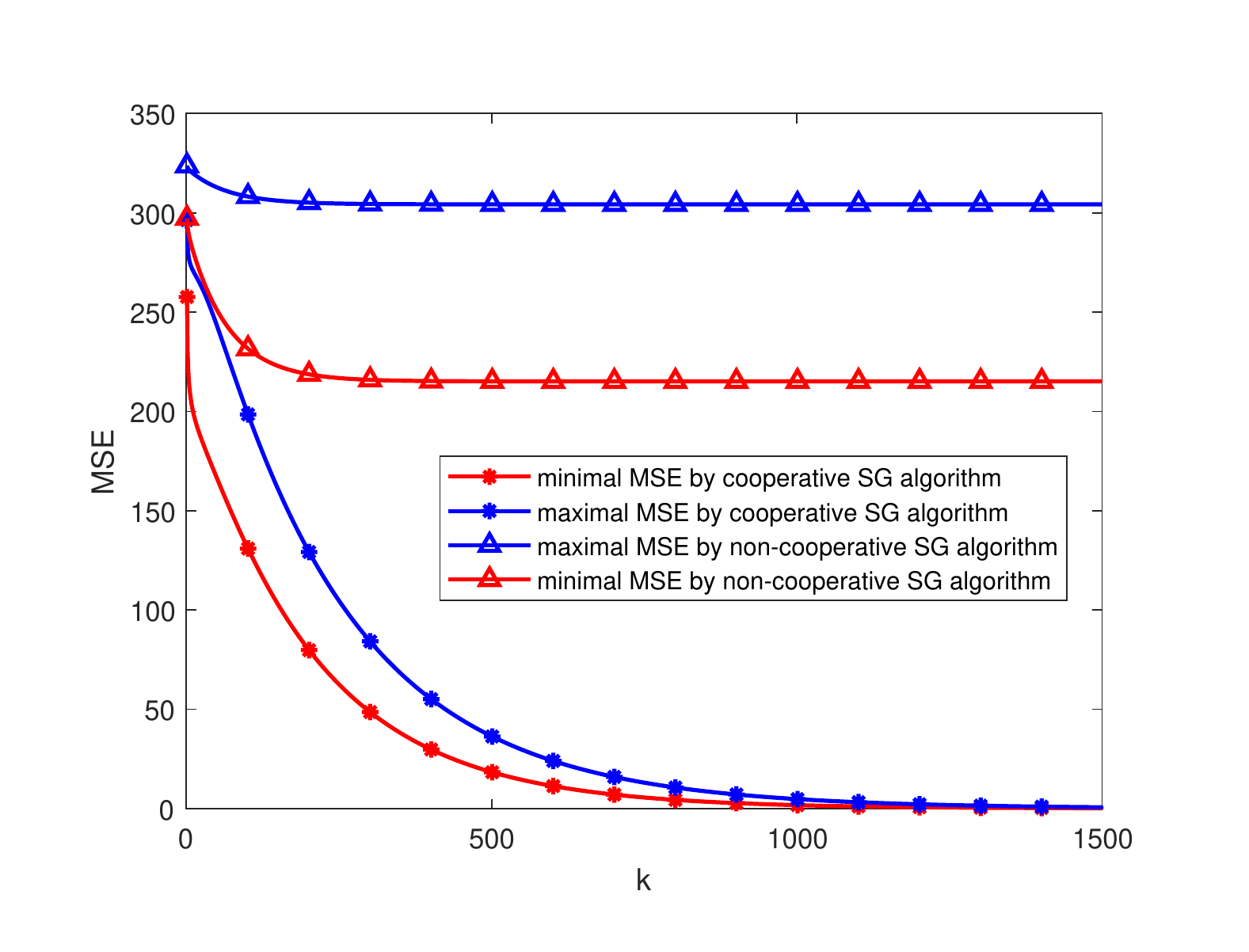}
\caption{ The maximal  and minimal MSEs of sensors by using non-cooperative SG algorithm and cooperative SG algorithm.}\label{f7}
\end{figure}

We can  verify that for each sensor $i (i=1,\cdots,28)$,
the regression signals $\bm\varphi_k^i($ generated by (\ref{phi}) can not satisfy the excitation condition (\ref{excitaion}), but they can cooperate to satisfy Assumption \ref{a2}. We repeat the simulation for $s = 500$ times with the same
initial states.
 For fixed step sizes $\mu=0.25$ and $\nu =0.7$, the simulation results are shown in Fig. \ref{f7}.
 We see that
if all the sensors use the
non-cooperative SG algorithm to estimate $\bm \theta$, the mean square error (MSE) of each sensor  cannot converge to zero,
while the MSE of each sensor in the distributed SG algorithm converges to zero. It is clear that the estimation task can be  fulfilled through exchanging information between sensors even though any individual sensor can not.

\section{Concluding Remarks}

This paper proposed a distributed SG algorithm based on the consensus strategy and the diffusion of the regression vectors to cooperatively estimate an unknown time-invariant parameter. We introduced a cooperative excitation condition, under which the almost sure convergence of the proposed algorithm can be guaranteed, and the convergence rate of the algorithm can be established. Compared with the existing results concerning the distributed estimation in the literature, our results are obtained without relying on the independency and stationarity assumptions, which makes it possible to apply our results to the feedback control systems. Furthermore, we found that the sensors can cooperate to finish the estimation task even though any individual can not. Many interesting problems  deserve to be further investigated, for example, the convergence of the distributed SG algorithm with correlated noise, the analysis of other distributed algorithms such as the distributed Kalman filter, and the combination of the distributed adaptive estimation with the distributed control.


%

\appendices
%
\section{Proof of Lemma \ref{l3} }\label{l3z}
\textbf{Proof.} By $\varepsilon<1$ we see that $2-\varepsilon-2\delta>1$.  Since $\bm R_k=\bm R_{k-1}+\bm\Phi^T_k\bm\Phi_k$, we obtain
\bna
tr(\bm R_k)=tr(\bm R_{k-1})+tr(\bm\Phi^T_k\bm\Phi_k). \label{rk}
\ena
Then by Lemma \ref{l2}, we have the following inequality,
\begin{gather*}
\sum^{\infty}_{k=1}\frac{\|\bm\Phi_k\|^2}{\|\bm R_k\|^{2-2\delta-\varepsilon}}\leq{n^{2-2\delta-\varepsilon}}\sum^{\infty}_{k=1}\frac{tr(\bm\Phi^T_k\bm\Phi_k)}{(tr(\bm R_k))^{2-2\delta-\varepsilon}}<\infty.
\end{gather*}
By the boundedness of the condition number of $\bm R_k$, we have $\|\bm R_k\|\|\bm R^{-1}_k\|\leq\gamma$ for all $k$. Using this and the above inequality, we have,
\ban
&&\sum^{\infty}_{k=1}E(\|\bm\Phi_k\bm R^{\delta-1}_k\bm\Xi^T_{k+1}\|^2|\mathscr{F}_k)\nonumber\\
&\leq &c_0\sum^{\infty}_{k=1}\|\bm\Phi_k\|^2\|\bm R^{\delta-1}_k\|^2\|\bm R_k\|^{\varepsilon}\\
&=&c_0\sum^{\infty}_{k=1}\|\bm\Phi_k\|^2\|\bm R^{-1}_k\|^{2-2\delta}\|\bm R_k\|^{\varepsilon}\nonumber\\
&\leq &c_0\gamma^{2-2\delta}\sum^{\infty}_{k=1}\frac{\|\bm\Phi_k\|^2}{\|\bm R_k\|^{2-2\delta-\varepsilon}}<\infty,~~{\rm a.s. }
\ean
where Assumption $\ref{a3}$ is used in the first inequality.
By the martingale convergence theorem it follows that as $k\rightarrow\infty$, $\sum^{k}_{j=1}\bm\Phi_j\bm R^{\delta-1}_j\bm\Xi^T_{j+1}$ converges a.s.. Hence  for  any $\eta>0$, if $k$ is large enough, then  we have $\|\widetilde{\bm S}_{k-1,\delta}\|<\eta$,  where $\widetilde{\bm S}_{k-1,\delta}\triangleq\sum^{\infty}_{j=k}\bm\Phi_j\bm R^{\delta-1}_j\bm\Xi^T_{j+1}$.

By the definition of $\bm \Phi_j$ and $\bm R_j$ in Table \ref{biao1}, we see that $\bm \Phi_j\bm R^{-\delta}_j=(\bm R^{-\delta}_j\otimes\bm I_m)\bm \Phi_j$ . Then summation by parts yields the following result,
\bna
&&\Big\|(\bm R^{\delta}_k\otimes\bm I_m)\widetilde{\bm S}_{k-1}\Big\|\nonumber\\
&=&\Big\|(\bm R^{\delta}_k\otimes\bm I_m)\sum^{\infty}_{j=k}(\bm R^{-\delta}_j\otimes\bm I_m)\bm \Phi_j\bm R^{\delta-1}_j\bm\Xi^T_{j+1}\Big\|\nonumber\\
&=&\Big\|(\bm R^{\delta}_k\otimes\bm I_m)\sum^{\infty}_{j=k}(\bm R^{-\delta}_j\otimes\bm I_m)({\widetilde{\bm S}_{j-1,\delta}}-{\widetilde{\bm S}_{j,\delta}})\Big\|\nonumber\\
&=&\Big\|{\widetilde{\bm S}_{k-1,\delta}}-(\bm R^{\delta}_k\otimes\bm I_m)\sum^{\infty}_{j=k}(( \bm R^{-\delta}_j- \bm R^{-\delta}_{j+1})\otimes\bm I_m){\widetilde{\bm S}_{j,\delta}}\Big\|\nonumber\\
&\leq&\eta+\eta\Big\|(\bm R^{\delta}_k\otimes\bm I_m)\sum^{\infty}_{j=k}\left(( \bm R^{-\delta}_j-\bm R^{-\delta}_{j+1})\otimes\bm I_m\right)\Big\| \nonumber\\
&\leq& 2\eta.\label{3.1}
\ena
Furthermore, by (\ref{3.1}) we have
\begin{align*}
\|\widetilde{\bm S}_{k-1}\|&=\|(\bm R^{-\delta}_k\otimes\bm I_m)(\bm R^{\delta}_k\otimes\bm I_m)\widetilde{\bm S}_{k-1}\|\\&\leq\|(\bm R^{-\delta}_k\otimes\bm I_m)\|\|(\bm R^{\delta}_k\otimes\bm I_m)\widetilde{\bm S}_{k-1}\|\\&\leq 2\eta\|(\bm R^{-1}_k\otimes\bm I_m)\|^{\delta}
=2\eta\|\bm R^{-1}_k\|^{\delta}\leq \frac{2\eta\gamma^{\delta}}{\|\bm R_k\|^{\delta}},
\end{align*}
where  $\|\bm R^{-1}_k\otimes\bm I_m\|=\|\bm R^{-1}_k\|$ is used. This completes the proof of the lemma. \hfill$\blacksquare$

\section{Proof of Theorem \ref{t2}}\label{t2z}
Before proving Theorem \ref{t2}, we first introduce two lemmas, whose proof can be found in \cite{c1}.
\begin{lemma}\label{l6}\cite{c1}
\ Suppose that $0\leq\bm C_k\leq\bm I_{p},~k\geq0$ with $\bm C_k\in \mathbb{R}^{p\times p}$.~Set~$\bm\Pi(k+1,j)=(\bm I-\bm C_k)\bm\Pi(k,j),~\bm\Pi(j,j)=\bm I,~\forall~k\geq j,$~ then we have
\begin{align}
\|\bm\Pi(M,k)\|\leq\left(1-\frac{\lambda_{\min}(\bm F_{kM})}{2(1+r^2_{kM})}\right)^{\frac{1}{2}}, \ \ \ M>k,
\end{align}
where~$\bm F_{kM}\triangleq\sum^{M-1}_{j=k}\bm C_j$, $r_{kM}\triangleq\sum^{M-1}_{j=k}\|\bm C_j\|$, and $\lambda_{\min}(\bm F_{kM})$~represents the minimum eigenvalue of~$\bm F_{kM}$.
\end{lemma}

\begin{lemma}\cite{c1}\label{l7}
\ Suppose that $0\leq\bm C_k\leq\bm I_{p},~k\geq0$ with $\bm C_k\in \mathbb{R}^{p\times p}$. Then $\bm \Pi(k,0)$ defined in Lemma \ref{l6} converges to 0 as $k\rightarrow\infty$, if there exists a sequence of monotonically increasing positive integers $\{t_k\}$ with $t_k\rightarrow\infty$ as $k\rightarrow \infty$,~such that
\begin{align}
\sum^{\infty}_{k=1}\frac{\lambda_{\min}(\bm F_k)}{1+\lambda^2_{\max}(\bm F_k)}=\infty,
\end{align}
where~$\bm F_k\triangleq\sum^{t_k-1}_{j=t_{k-1}}\bm C_j$.
\end{lemma}
\textbf{Proof of Theorem \ref{t2}.}

\textbf{Proof.}
By Lemma \ref{l7}, we just need to show that there exists an integer sequence $\{t_k\}$  with $t_k\rightarrow\infty$ as $k\rightarrow \infty$, such that
\bna
\sum^{\infty}_{k=1}\frac{\lambda_{\min}\left(\sum^{t_k-1}_{j=t_{k-1}}\mu\bm G_j\right)}{1+\lambda^2_{\max}\left(\sum^{t_k-1}_{j=t_{k-1}}\mu\bm G_j\right)}=\infty.\label{infty}
\ena
In the following, we will show this by three steps.

\textbf{Step 1: Construction of the integer sequence $\{t_k\}$.}

Without loss of generality, the constant $K_0$ in Assumption \ref{a2} can be taken to satisfy the inequality $\log tr(\bm R_{K_0})\geq1$  since $\|\bm R_k\|\rightarrow\infty$ as $k\rightarrow\infty$.
 Thus, by the definition of $d_k$ and $\hat{g}(t)$ in Lemma \ref{l5}, we have for $t\geq0$,
\begin{align}
t\leq d_{\hat{g}(t)+1}\leq d_{\hat{g}(t)}+1\leq t+1.\label{3.20}
\end{align}

We first prove the following result,
\begin{gather}
\hat{g}(t)\rightarrow\infty,~t\rightarrow\infty.\label{72}
\end{gather}
By the boundedness of the condition number of $\bm R_k$, we have for large $k$
\begin{align}
\frac{\|\bm R_k\|}{\|\bm R_{k-1}\|}&=\frac{\max_ir^i_k}{\max_ir^i_{k-1}}=\frac{\max_ir^i_k}{\min_ir^i_{k}}\cdot\frac{\min_ir^i_k}{\max_ir^i_{k-1}}\nonumber\\
&\leq\frac{\max_ir^i_k}{\min_ir^i_{k}}\cdot\frac{r^{i_1}_k}{r^{i_1}_{k-1}}\leq \gamma r^*.\nonumber
\end{align}
Hence  we have
\begin{align}
\frac{tr\bm R_k}{tr\bm R_{k-1}}\leq\frac{n\|\bm R_k\|}{\|\bm R_{k-1}\|}\leq n\gamma r^*.\label{g1}
\end{align}
Then by (\ref{rk}) and (\ref{g1}), we have for any $k\geq K_0$,
\begin{align}
d_k&=\sum^{k-1}_{j=K_0}\frac{tr\bm R_j-tr\bm R_{j-1}}{tr(\bm R_{j})(\log tr(\bm R_{j-1}))^{\frac{1}{3}}}\nonumber\\
&\geq\frac{1}{n\gamma r^*}\sum^{k-1}_{j=K_0}\frac{tr\bm R_j-tr\bm R_{j-1}}{tr(\bm R_{j-1})(\log tr(\bm R_{j-1}))^{\frac{1}{3}}}\nonumber\\
&\geq\frac{1}{n\gamma r^*}\sum^{k-1}_{j=K_0}\int^{tr\bm R_j}_{tr\bm R_{j-1}}\frac{dx}{x(\log x)^{\frac{1}{3}}}\nonumber\\
&=\frac{3}{2n\gamma r^*}\left(\log^{\frac{2}{3}}tr\bm R_{k-1}-\log^{\frac{2}{3}}tr\bm R_{K_0-1}\right).\label{75}
\end{align}
Since $\|\bm R_k\|\xrightarrow[]{k\rightarrow\infty}\infty$, then we have $d_k\rightarrow\infty$ as $k\rightarrow\infty$. By the definition of $\hat{g}(t)$, we further have  $\hat{g}(t)<\infty$ for all $t>0$ and $\hat{g}(t)\rightarrow\infty$ as $t\rightarrow\infty.$

By (\ref{72}) and $\|\bm R_j\|\xrightarrow[]{j\rightarrow\infty}\infty$, we see that there exists sufficiently large $N_1$ such that for $j\geq \hat{g}(N_1)$,
\bna\label{3.24}
 \frac{n(\log tr\bm R_j)^{\frac{1}{3}}}{tr\bm R_j}\leq\frac{1}{2N},
\ena
where $N$ is defined in Assumption \ref{a2}.
The integer sequence $\{t_k\}$ can be taken as
\bna\label{alpha}
t_k=\hat{g}(N_1+k\alpha), \ \ \alpha=2Nm(n+2)+1.\ena
Then by (\ref{72}), we have $t_k\rightarrow\infty$ as $k\rightarrow\infty$.

\textbf{Step 2: Estimation of $\lambda_{\min}\left(\sum^{t_k-1}_{j=t_{k-1}}\mu\bm G_j\right)$.}

By the properties of $r_j^i$, we have
\bna
& &\sum^n_{i=1}\sum^{t_k-1}_{j=t_{k-1}}\frac{\bm \varphi^i_j{\bm\varphi^i_j}^T}{r^i_j}
\geq\sum^n_{i=1}\sum^{t_k}_{j=t_{k-1}}\frac{\bm \varphi^i_j{\bm\varphi^i_j}^T}{r^i_j}-n\bm I_m\nonumber\\
&=&\sum^{t_k}_{j=t_{k-1}}\sum^n_{i=1}\frac{1}{r^i_j}\left(\sum^j_{l=1}\bm \varphi^i_l{\bm\varphi^i_l}^T-\sum^{j-1}_{l=1}\bm \varphi^i_l{\bm\varphi^i_l}^T\right)-n\bm I_m\nonumber\\
&\geq&\sum^{t_k}_{j=t_{k-1}}\frac{1}{tr\bm R_j}\sum^n_{i=1}\left(\sum^j_{l=1}\bm \varphi^i_l{\bm\varphi^i_l}^T-\sum^{j-1}_{l=1}\bm \varphi^i_l{\bm\varphi^i_l}^T\right)-n\bm I_m\nonumber\\
&=&\sum^{t_k+1}_{j=t_{k-1}+1}\frac{1}{tr\bm R_{j-1}}\sum^n_{i=1}\sum^{j-1}_{l=1}\bm \varphi^i_l{\bm\varphi^i_l}^T-n\bm I_m\nonumber\\
&&-\sum^{t_k}_{j=t_{k-1}}\frac{1}{tr\bm R_{j}}\sum^n_{i=1}\sum^{j-1}_{l=1}\bm \varphi^i_l{\bm\varphi^i_l}^T.\label{x2}
\ena
By  the definition of $\lambda^{(k)}_{\min}$ in Assumption \ref{a2} and (\ref{x2}), we obtain
\bna
& &\sum^n_{i=1}\sum^{t_k-1}_{j=t_{k-1}}\frac{\bm \varphi^i_j{\bm\varphi^i_j}^T}{r^i_j}\nonumber\\
&\geq&\sum^{t_k}_{j=t_{k-1}+1}\left(\frac{1}{tr\bm R_{j-1}}-\frac{1}{tr\bm R_{j}}\right)\sum^n_{i=1}\sum^{j-1}_{l=1}\bm \varphi^i_l{\bm\varphi^i_l}^T-n\bm I_m\nonumber\\
&&+\frac{1}{tr\bm R_{t_k}}\sum^n_{i=1}\sum^{t_k}_{l=1}\bm \varphi^i_l{\bm\varphi^i_l}^T
-\frac{1}{tr\bm R_{t_{k-1}}}\sum^n_{i=1}\sum^{t_{k-1}}_{l=1}\bm \varphi^i_l{\bm\varphi^i_l}^T\nonumber\\
&\geq&\sum^{t_k}_{j=t_{k-1}+1}\left(\frac{tr(\bm\Phi^T_j\bm\Phi_j)}{tr\bm R_{j}tr\bm R_{j-1}}\right)\sum^n_{i=1}\sum^{j-1}_{l=1}\bm \varphi^i_l{\bm\varphi^i_l}^T-(n+1)\bm I_m\nonumber\\
&\geq&\sum^{t_k}_{j=t_{k-1}+1}\left(\lambda^{(j-1)}_{\min}-\frac{n}{m}\right)\cdot\left(\frac{tr(\bm\Phi^T_j\bm\Phi_j)}{tr\bm R_{j}tr\bm R_{j-1}}\right)\bm I_m\nonumber\\
& &\hskip 4.5cm-(n+1)\bm I_m.
\ena
Combining this with Assumption \ref{a2}, we have
\bna\label{88}
& &\sum^n_{i=1}\sum^{t_k-1}_{j=t_{k-1}}\frac{\bm \varphi^i_j{\bm\varphi^i_j}^T}{r^i_j}+(n+1)\bm I_m\nonumber\\
&\geq&\sum^{t_k}_{j=t_{k-1}+1}\Big(\frac{\lambda^{(j-1)}_{\max}}{N(\log tr\bm R_{j-1})^{\frac{1}{3}}}-\frac{n}{m}\Big)\Big(\frac{tr(\bm\Phi^T_j\bm\Phi_j)}{tr\bm R_{j}tr\bm R_{j-1}}\Big)\bm I_m\nonumber\\
&\geq&\frac{1}{m}\sum^{t_k}_{j=t_{k-1}+1}\Big(\frac{tr\bm R_{j-1}}{N(\log tr\bm R_{j-1})^{\frac{1}{3}}}-n\Big)\Big(\frac{tr(\bm\Phi^T_j\bm\Phi_j)}{tr\bm R_{j}tr\bm R_{j-1}}\Big)\bm I_m\nonumber\\
&=&\frac{1}{m}\sum^{t_k}_{j=t_{k-1}+1}\left(\frac{1}{N}-\frac{n(\log tr\bm R_{j-1})^{\frac{1}{3}}}{tr\bm R_{j-1}}\right)\nonumber\\
&&\hskip3cm\cdot\left(\frac{tr(\bm\Phi^T_j\bm\Phi_j)}{tr\bm R_{j}(\log tr\bm R_{j-1})^{\frac{1}{3}}}\right)\bm I_m\nonumber\\
&\geq&\frac{1}{2Nm}\sum^{t_k}_{j=t_{k-1}+1}\left(\frac{tr(\bm\Phi^T_j\bm\Phi_j)}{tr\bm R_{j}(\log tr\bm R_{j-1})^{\frac{1}{3}}}\right)\bm I_m\nonumber,\ena
where \dref{3.24} is used in the last inequality.
Hence by the definition of $d_{k}$ in Lemma \ref{l5}, we can obtain
\ban
&&\sum^n_{i=1}\sum^{t_k-1}_{j=t_{k-1}}\frac{\bm \varphi^i_j{\bm\varphi^i_j}^T}{r^i_j}\\
&\geq&\frac{1}{2Nm}(d_{t_k+1}-d_{t_{k-1}+1})\bm I_m-(n+1)\bm I_m.
\ean
Furthermore,  by Lemma \ref{l5},  (\ref{3.20})  and (\ref{alpha}), we have
\bna\label{89}
&&\lambda_{\min}\left(\sum^{t_k-1}_{j=t_{k-1}}\bm G_j\right)\geq\sigma \lambda_{\min}\left(\sum^n_{i=1}\sum^{t_k-1}_{j=t_{k-1}}\frac{\bm \varphi^i_j{\bm\varphi^i_j}^T}{r^i_j}\right)\nonumber\\
&\geq&\frac{\sigma}{2Nm}(d_{\hat{g}(N_1+k\alpha)+1}-d_{\hat{g}(N_1+(k-1)\alpha+1)})-\sigma(n+1)\nonumber\\
&\geq&\frac{\sigma}{2Nm}(N_1+k\alpha-(N_1+(k-1)\alpha+1))-\sigma(n+1)\nonumber\\
&=&\sigma\left(\frac{\alpha-1}{2Nm}-(n+1)\right)=\sigma.\label{neww2}
\ena

\textbf{Step 3: Estimation of  $\lambda_{\max}\left(\sum^{t_k-1}_{j=t_{k-1}}\mu\bm G_j\right)$.}

By the basic properties of the trace and the Euclidean norm of the matrix, we have
\bna
&&\sum^{t_k-1}_{j=t_{k-1}}\|\bm A_j\|\nonumber\\
&\leq&\sum^{t_k-1}_{j=t_{k-1}}tr(\bm R^{-1}_j\bm\Phi^T_j\bm\Phi_j)
\leq\sum^{t_k-1}_{j=t_{k-1}}tr(\bm\Phi^T_j\bm\Phi_j)\|\bm R^{-1}_j\|\nonumber\\
&\leq&\gamma\sum^{t_k-1}_{j=t_{k-1}}\frac{tr(\bm\Phi^T_j\bm\Phi_j)}{\|\bm R_j\|}
\leq n\gamma\sum^{t_k-1}_{j=t_{k-1}}\frac{tr(\bm\Phi^T_j\bm\Phi_j)}{tr\bm R_j}\nonumber\\
&\leq& n\gamma(\log tr\bm R_{t_k-1})^{\frac{1}{3}}\sum^{t_k-1}_{j=t_{k-1}}\frac{tr(\bm\Phi^T_j\bm\Phi_j)}{tr\bm R_j(\log tr\bm R_{j-1})^{\frac{1}{3}}}.\label{3.28}
\ena
By the definition of $d_{\hat{g}(t)}$, (\ref{3.20}), and (\ref{3.28}), we have
\ban
&&\sum^{t_k-1}_{j=t_{k-1}}\|\bm A_j\|
\leq n\gamma(\log tr\bm R_{t_k-1})^{\frac{1}{3}}(d_{t_k}-d_{t_{k-1}})\\
&= &n\gamma(\log tr\bm R_{t_k-1})^{\frac{1}{3}}(d_{\hat{g}(N_1+k\alpha)}-d_{\hat{g}(N_1+(k-1)\alpha)})\\
&\leq &n\gamma(\alpha+1)(\log tr\bm R_{t_k-1})^{\frac{1}{3}}.
\ean
Then combining this with \dref{1}, we have
\bna
&&\lambda_{\max}\left(\sum^{t_k-1}_{j=t_{k-1}}\bm G_j\right)\nonumber\\
&\leq&\sum^{t_k-1}_{j=t_{k-1}}\|\bm G_j\|
\leq(1+4\nu)\sum^{t_k-1}_{j=t_{k-1}}\|\bm A_j\|\nonumber\\
&\leq& n\gamma(1+4\nu)(\alpha+1)(\log tr\bm R_{t_k-1})^{\frac{1}{3}}.\label{3.29}
\ena
According to (\ref{3.20}), (\ref{75}) and (\ref{alpha}), we can see that
\bna
&&N_1+k\alpha\geq d_{\hat{g}(N_1+k\alpha)}\nonumber\\
&\geq&\frac{3}{2n\gamma r^*}\left(\log^{\frac{2}{3}}tr\bm R_{t_k-1}-\log^{\frac{2}{3}}tr\bm R_{K_0-1}\right).\label{3.30}
\ena
By (\ref{3.29}) and (\ref{3.30}), we have
\bna
&&\lambda^2_{\max}\left(\sum^{t_k-1}_{j=t_{k-1}}\bm G_j\right)\nonumber\\
&\leq& n^2\gamma^2(1+4\nu)^2(\alpha+1)^2\nonumber\\
&&\cdot\left(\frac{2n\gamma r^*}{3}(N_1+k\alpha)+\log^{\frac{2}{3}}tr\bm R_{K_0-1}\right)\nonumber\\
&\triangleq& p_k=O(k).\label{neww3}
\ena
Combining this with $(\ref{89})$ yields
\bna
&&\sum^{\infty}_{k=1}\frac{\lambda_{\min}\left(\sum^{t_k-1}_{j=t_{k-1}}\mu\bm G_j\right)}{1+\lambda^2_{\max}\left(\sum^{t_k-1}_{j=t_{k-1}}\mu\bm G_j\right)}\nonumber\\
&\geq&\sum^{\infty}_{k=1}\frac{\mu\sigma}{1+\mu^2p_k}
=\infty.\label{100}
\ena
According to Lemma \ref{l7}, we can see that ${\bm\Psi}(k,0)\rightarrow0$ as $k\rightarrow\infty$. This completes the proof of the theorem. \hfill$\blacksquare$

\section{Proof of Lemma \ref{l9}}\label{s4}
\textbf{Proof.}
By the definition of ${\bm\Psi}(k,j)$ in (\ref{11}), it is clear that we have $(i)$.

By Lemma \ref{l8}, we have
\bna
&&det{\bm\Psi}(k+1,0)\nonumber\\
&=&\prod^k_{j=0}det(\bm I_{mn}-\mu\bm G_j)\geq\prod^k_{j=0}(det(\bm I_{mn}-\bm A_j))^{\tau_1}\nonumber\\
&=&(det(\bm I_{mn}-\bm A_0))^{\tau_1}\left(\prod^n_{i=1}\prod^k_{j=1}\frac{r^i_{j-1}}{r^i_j}\right)^{\tau_1}\nonumber\\
&=&(det(\bm I_{mn}-\bm A_0))^{\tau_1}\prod^n_{i=1}\frac{1}{(r^i_k)^{\tau_1}}\nonumber\\
&=&\frac{1}{det(\bm R^{\tau_1}_k)}\prod^n_{i=1}(1-\|\bm\varphi^i_0\|^2)^{\tau_1}\nonumber\\
&\geq&\frac{1}{\|\bm R_k\|^{n\tau_1}}\prod^n_{i=1}(1-\|\bm\varphi^i_0\|^2)^{\tau_1}.\label{122}
\ena
Therefore, we have
\ban
&&\left(\frac{1}{\|\bm R_k\|^{n\tau_1}}\prod^n_{i=1}(1-\|\bm\varphi^i_0\|^2)^{\tau_1}\right)^2\\
&\leq& det({\bm\Psi}(k+1,0){\bm\Psi^T}(k+1,0))\leq\|{\bm\Psi}(k+1,0)\|^{2mn}.
\ean
Since the initial value $\bm\varphi^i_0$ can be arbitrarily selected, without loss of generality we suppose $\|\bm\varphi^i_0\|\neq1$ for all $i\in\{1,...,n\}$. It is clear that $(ii)$ of the lemma holds.

By Lemma \ref{l8}, we have
\bna
&&\|{\bm\Psi}(k,j+1)\|\leq\|{\bm\Psi}(k,0)\|\|\bm\Psi^{-1}(j+1,0)\|\nonumber\\
&\leq&\|{\bm\Psi}(k,0)\|\prod^{j+1}_{p=1}\|(\bm I_{mn}-\mu\bm G_{p-1})^{-1}\|\nonumber\\
&\leq&\|{\bm\Psi}(k,0)\|\prod^{j+1}_{p=1}det\left((\bm I_{mn}-\mu\bm G_{p-1})^{-1}\right)\nonumber\\
&\leq&\|{\bm\Psi}(k,0)\|\prod^{j+1}_{p=1}\frac{1}{\left(det(\bm I_{mn}-\bm A_{p-1})\right)^{\tau_1}}\nonumber\\
&=&\|{\bm\Psi}(k,0)\|\left(\frac{1}{det(\bm I_{mn}-\bm A_0)}\right)^{\tau_1}\prod^n_{i=1}\prod^{j+1}_{p=2}\left(\frac{r^i_{p-1}}{r^i_{p-2}}\right)^{\tau_1}\nonumber\\
&=&\|{\bm\Psi}(k,0)\|\cdot\left(\frac{1}{det(\bm I_{mn}-\bm A_0)}\right)^{\tau_1}\left(\prod^n_{i=1}r^i_j\right)^{\tau_1}\nonumber\\
&\leq&\left(\frac{1}{det(\bm I_{mn}-\bm A_0)}\right)^{\tau_1}\cdot\|{\bm\Psi}(k,0)\|\cdot\|\bm R_j\|^{n\tau_1}.
\ena
Hence $(iii)$ of the lemma can be proved.

Now we will prove $(iv)$.
\bna
&&\sum^{\infty}_{j=M+1}\frac{\|\bm\Phi_j\|^2}{\|\bm R_j\|^{1+{\varsigma}}}
\leq\sum^{\infty}_{j=M+1}\frac{tr(\bm\Phi^T_j\bm\Phi_j)}{\frac{1}{n^{1+{\varsigma}}}(tr\bm R_j)^{1+{\varsigma}}}\nonumber\\
&=&n^{1+{\varsigma}}\sum^{\infty}_{j=M+1}\int^{tr\bm R_j}_{tr\bm R_{j-1}}\frac{1}{(tr\bm R_j)^{1+{\varsigma}}}~dt\nonumber\\
&\leq& n^{1+{\varsigma}}\sum^{\infty}_{j=M+1}\int^{tr\bm R_j}_{tr\bm R_{j-1}}\frac{1}{t^{1+{\varsigma}}}~dt\nonumber\\
&\leq& n^{1+{\varsigma}}\int^{\infty}_{tr\bm R_M}\frac{1}{t^{1+{\varsigma}}}~dt\nonumber\\
&=&\frac{n^{1+{\varsigma}}}{{\varsigma}}\frac{1}{(tr\bm R_M)^{{\varsigma}}}
\leq\frac{n^{1+{\varsigma}}}{{\varsigma}}\frac{1}{\|\bm R_M\|^{{\varsigma}}}.\nonumber
\ena
This completes the proof of the lemma. \hfill$\blacksquare$

\section{Proof of Theorem \ref{t4}}\label{tt4}
\textbf{Proof.}
By (\ref{neww2}) and (\ref{neww3}) in Appendix \ref{t2z}, there exists a constant $c^*>0$ such that
\ban
 \frac{\lambda_{\min}\left(\sum^{t_k-1}_{j=t_{k-1}}\mu\bm G_j\right)}{1+\lambda^2_{\max}\left(\sum^{t_k-1}_{j=t_{k-1}}\mu\bm G_j\right)}\geq\frac{c^*}{1+k},
 \ean
where $t_k$ is defined in (\ref{alpha}) of Appendix \ref{t2z}. In fact, since $\mu(1+4\nu)<1$, we can take
\begin{align*}
 &c^*\\
 =&\frac{\mu\sigma}{1+[n\gamma(\alpha+1)]^2\max\{\frac{2n\gamma r^* \alpha}{3},\frac{2n\gamma r^* N_1}{3}+\log^{\frac{2}{3}}tr\bm R_{K_0-1}\}}
 \end{align*}
  with $\alpha=2Nm(n+2)+1$, $N_1\propto n$, where $\sigma=\frac{l^2_{m+1}a\nu}{2n+l^2_{m+1}a\nu n}$, $\gamma$ and $r^*$  are respectively defined in Lemma  \ref{l3}
 and Theorem \ref{t2}.

By Lemma \ref{l6}, we have
\bna
&&\|{\bm\Psi}(t_k,t_{k-1})\|\nonumber\\
&\leq&\left(1-\frac{\lambda_{\min}\left(\sum^{t_k-1}_{j=t_{k-1}}\mu\bm G_j\right)}{2\left(1+\left(\sum^{t_k-1}_{j=t_{k-1}}\|\mu\bm G_j\|\right)^2\right)}\right)^{\frac{1}{2}}\nonumber\\
&\leq&\left(1-\frac{\lambda_{\min}\left(\sum^{t_k-1}_{j=t_{k-1}}\mu\bm G_j\right)}{2\left(1+m^2n^2\lambda^2_{\max}\left(\sum^{t_k-1}_{j=t_{k-1}}\mu\bm G_j\right)\right)}\right)^{\frac{1}{2}}\nonumber\\
&\leq&\left(1-\frac{\lambda_{\min}\left(\sum^{t_k-1}_{j=t_{k-1}}\mu\bm G_j\right)}{2m^2n^2\left(1+\lambda^2_{\max}\left(\sum^{t_k-1}_{j=t_{k-1}}\mu\bm G_j\right)\right)}\right)^{\frac{1}{2}}\nonumber\\
&\leq&\left(1-\frac{\bar{c}}{1+k}\right)^{\frac{1}{2}}  ~~~~~\left(\bar{c}=\frac{c^*}{2m^2n^2}\right).
\ena

By the definition of $t_k$ in (\ref{alpha}), we have the following estimate about $\|{\bm\Psi}(\hat{g}(N_1+j\alpha),0)\|$ with $N_1$ and $\alpha$  defined  in Appendix \ref{t2z}.
\bna
&&\|{\bm\Psi}(\hat{g}(N_1+j\alpha),0)\|\nonumber\\
&\leq&\prod^j_{l=1}\|{\bm\Psi}(\hat{g}(N_1+l\alpha),\hat{g}(N_1+(l-1)\alpha)\|\nonumber\\
&&\cdot\|{\bm\Psi}(\hat{g}(N_1),0)\|\nonumber\\
&\leq&\prod^j_{l=1}\left(1-\frac{\bar{c}}{1+l}\right)^{\frac{1}{2}}
\leq\prod^j_{l=1}e^{-\frac{\bar{c}}{2(1+l)}}\nonumber\\
&=&e^{-\sum^j_{l=1}\frac{\bar{c}}{2(1+l)}}\leq e^{-\frac{\bar{c}}{2}\cdot\log({\frac{j+2}{2}})}=\left(\frac{j+2}{2}\right)^{-\frac{\bar{c}}{2}},\label{138}
\ena
where the inequalities $1-x\leq e^{-x} $ for all $x\geq0$ and $\sum^k_{j=1}\frac{1}{1+j}\geq\log^{\frac{k+2}{2}}$ for all $~k\geq1$ are used.

Since $\hat{g}(t)\rightarrow\infty$ as $t\rightarrow\infty$ in (\ref{72}), then for any $k\geq \hat{g}(N_1+\alpha)$, there exists $j\geq1$ such that
\begin{align*}
\hat{g}(N_1+j\alpha)\leq k\leq \hat{g}(N_1+(j+1)\alpha).
\end{align*}
 By  the monotonicity of $d_k$ and (\ref{3.20}), we have
\ban
d_k\leq d_{\hat{g}(N_1+(j+1)\alpha)}\leq N_1+(j+1)\alpha.
\ean
Thus,  $j\geq\frac{d_k-N_1-\alpha}{\alpha}$. According to  $(\ref{138})$, we obtain
\bna
&&\|{\bm\Psi}(k,0)\|\nonumber\\
&\leq&\|{\bm\Psi}(k,\hat{g}(N_1+j\alpha))\|\cdot\|{\bm\Psi}(\hat{g}{(N_1+j\alpha)},0)\|\nonumber\\
&\leq&\|{\bm\Psi}(\hat{g}(N_1+j\alpha),0)\|\leq\left(\frac{d_k-N_1+\alpha}{2\alpha}\right)^{-\frac{\bar{c}}{2}}.\label{x1}
\ena
Combining (\ref{75}) with (\ref{x1}), we have for large $k$
\ban
\|{\bm\Psi}(k,0)\|&=&O(\log tr\bm R_{k-1})^{-\frac{\bar{c}}{3}}\nonumber\\
&=&O(\log tr\bm R_{k})^{-\frac{\bar{c}}{3}}.
\ean
Hence by Lemma \ref{t3}, we have
$
\|\widetilde{\bm\Theta}_{k}\|=O((\log\|\bm R_k\|)^{-\delta_1})$ with $\delta_1=\frac{\delta \bar{c}}{3n\tau_1(1+\delta)}=\frac{\delta c^*}{6m^3n^4(1+\delta)}.
$ \hfill$\blacksquare$

%
%

\ifCLASSOPTIONcaptionsoff
  \newpage
\fi

\end{document}